# OSTSC: Over Sampling for Time Series Classification in R

**Matthew Dixon**
Stuart School of Business,
Illinois Institute of Technology

**Diego Klabjan**
Department of Industrial Engineering and Management Sciences,
Northwestern University

**Lan Wei**
Department of Computer Science,
Illinois Institute of Technology

**Abstract**

The **OSTSC** package is a powerful oversampling approach for classifying univariant, but multinomial time series data in R. This article provides a brief overview of the oversampling methodology implemented by the package. A tutorial of the **OSTSC** package is provided. We begin by providing three test cases for the user to quickly validate the functionality in the package. To demonstrate the performance impact of **OSTSC**, we then provide two medium size imbalanced time series datasets. Each example applies a TensorFlow implementation of a Long Short-Term Memory (LSTM) classifier - a type of a Recurrent Neural Network (RNN) classifier - to imbalanced time series. The classifier performance is compared with and without oversampling. Finally, larger versions of these two datasets are evaluated to demonstrate the scalability of the package. The examples demonstrate that the **OSTSC** package improves the performance of RNN classifiers applied to highly imbalanced time series data. In particular, **OSTSC** is observed to increase the AUC of LSTM from 0.543 to 0.784 on a high frequency trading dataset consisting of 30,000 time series observations.

*Keywords*: Time Series Oversampling, Classification, Outlier Detection, R.

## 1. Introduction

A significant number of learning problems involve the accurate classification of rare events or outliers from time series data. For example, the detection of a flash crash, rogue trading, or heart arrhythmia from an electrocardiogram. Due to the rarity of these events, machine learning classifiers for detecting these events may be biased towards avoiding false positives. This is because any potential for false positives is greatly exaggerated by the number of negative samples in the data set.

Class imbalance problems are most easily addressed by treating the observations as conditionally independent. Then standard statistical techniques, such as oversampling the minority class or undersampling the majority class, or both, are applicable. (More 2016) compared a batch of resampling techniques' classification performances on imbalanced datasets. Besides the conventional resampling approaches, More showed how ensemble methods retain as much original information from the majority class as possible when performing undersampling. Ensemble methods perform well and have gained popularity in the data mining literature. (Dubey, Zhou, Wang, Thompson, and Ye 2014) studied an ensemble system of feature selection and data sampling from an imbalanced Alzheimer's Disease Neuroimaging Initiative dataset.

However the imbalanced time series classification problem is more complex when the time dimension needs to be accounted for. Not only is the assumption that the observations are conditionally independent too strong, but also the predictors may be cross-correlated too. The sample correlation structure may weaken or be entirely lost under application of the conventional resampling approaches described above.

There are two existing research directions for imbalanced time series classification. One is to preserve the covariance structure during oversampling proposed by (Cao, Li, Woon, and Ng 2011). Another is to conduct undersampling with various learning algorithms, proposed by (Liang and Zhang 2012). Both approaches are limited to binary classification and do not consider the more general problem of multi-classification.

A key assertation by (Cao, Tan, and Pang 2014) is that a time series sampling scheme should preserve the covariance structure. When the observations are conditionally dependent, this approach has been shown to outperform other sampling approaches such as undersampling the majority class, oversampling the minority class, and SMOTE. Our R package 'Over Sampling for Time Series Classification' (OSTSC) is built on this idea. OSTSC first implements Enhanced Structure Preserving Oversampling (EPSO) of the minority class. It then uses a nearest neighbor method from the SMOTE family to generate synthetic positives. Specifically, it uses an Adaptive Synthetic Sampling Approach for Imbalanced Learning (ADASYN). Note that other packages such as (Siriseriwan 2017) already implement SMOTE sampling techniques, including ADASYN. However an implementation of ADASYN has been provided in OSTSC for compatibility with the format required for use with EPSO and TensorFlow.

For examining the performance of oversampling for times series classification, RNNs are preferred (Graves 2013). Recently (Dixon 2017) applied RNNs to imbalanced times series data used in high frequency trading. The RNN classifier predicts a price-flip in the limit order book based on a sequence of limit order book depths and market orders. The approach uses standard under-sampling of the majority class to improve the classifier performance. OSTSC provides a uni-variate sample of this data and demonstrates the application of EPSO and ADASYN to improve the performance of the RNN. The RNN is implemented in 'TensorFlow' (Abadi, Barham, Chen, Chen, Davis, Dean *et al.* 2016) and made available in R by using a wrapper for 'Keras' (Allaire and Chollet 2017), a high-level API for 'TensorFlow'.

The current version of the package currently only supports univariant classification of time series. The extension to multi-features requires tensor computations which are not implemented here.

## 2. Overview

This article provides a brief description of the sampling methodologies implemented. We introduce the **OSTSC** package and illustrate its application using various examples. For validation purposes only, we first apply the OSTSC package to three small built-in toy datasets. These datasets are not sufficiently large to demonstrate the methodology. However, they can be used to quickly verify that the OSTSC function generates a balanced dataset.

For demonstrating the effect of OSTSC on LSTM performance, we provide two medium size datasets that can be computed with moderate computation. Finally, to demonstrate scalability, we evaluate OSTSC on two larger datasets. The reader is advised that the total amount of computation in this case is significant. We would therefore expect a user to test the OSTSC functionality on the small or medium size datasets, but reserve running the larger dataset examples on a higher performance machine. The medium and large datasets are not built-in to keep the package size within 5MB.

## 3. Background

ESPO is used to generate a large percentage of the synthetic minority samples from univariate labeled time series under the modeling assumption that the predictors are Gaussian. EPSO estimates the covariance structure of the minority-class samples and applies a spectral filter to reduce noise. ADASYN is a nearest neighbor interpolation approach which is subsequently applied to the EPSO samples (Cao, Li, Woon, and Ng 2013).

More formally, given the time series of positive labeled predictors $P = \left\{x_{11}, x_{12}, ..., x_{1|P|}\right\}$ and the negative time series $N = \left\{x_{01}, x_{02}, ..., x_{0|N|}\right\}$, where $|N| \gg |P|$, $x_{ij} \in R^{n\times 1}$, the new samples are generated by the following steps:

1. Removal of the Common Null Space: Let $q_{ij} = L_s^T x_{ij}$ represent $x_{ij}$ in a lower-dimensional signal space, where $L_s$ consists of eigenvectors in the signal space.

2. ESPO: Let

$$W_p = \frac{1}{|P|}\sum_{j=1}^{|P|}(q_{1j} - \bar{q}_1)(q_{1j} - \bar{q}_1)^T.$$

   and let $\hat{D} = V^T W_p V$ be the eigendecomposition with the diagonal matrix $\hat{D}$ of regularized eigenvalues $\left\{\hat{d}_1, ..., \hat{d}_n\right\}$, sorted in descending order, and with orthogonal eigenvector matrix $V$. Then we generate a synthetic positive sample from

$$b = \hat{D}^{1/2} V^T z + \bar{q}_1.$$

   $z$ is drawn from a zero-mean mixed Gaussian distribution and $\bar{q}_1$ is the corresponding positive-class mean vector. The oversampling is repeated until all $(|N| - |P|)r$ required synthetic samples are generated, where $r \in [0, 1]$ is the integration percentage of synthetic samples contributed by ESPO, which is chosen empirically. The remaining $(1-r)$ percentage of the samples are generated by the interpolation procedure described next.

3. ADASYN: Given the transformed positive data $P_t = \{q_{1i}\}$ and negative data $N_t = \{q_{0j}\}$, each sample $q_{1i}$ is replicated $\Gamma_i = |S_{i:k-NN} \bigcap N_t| / Z$ times, where $S_{i:k-NN}$ is this sample's kNN in the entire dataset, $Z$ is a normalization factor so that $\sum_{i=1}^{|P_t|} \Gamma_i = 1$.

   See (Cao *et al.* 2013) for further technical details of this approach.

## 4. Functionality

The package imports packages **parallel** (R Core Team 2017), **doParallel** (Microsoft Corporation and Weston 2017a), **doSNOW** (Microsoft Corporation and Weston 2017b) and **foreach** (Revolution Analytics and Weston 2015) for multi-threaded execution on shared memory architectures. Parallel execution is strongly suggested for datasets consisting of at least 30,000 observations. OSTSC also imports `mvrnorm()` from the **MASS** package (Venables and Ripley 2002) to generate random vectors from the multivariate normal distribution, and `rdist()` from the **fields** package (Douglas Nychka, Reinhard Furrer, John Paige, and Stephan Sain 2015) in order to calculate the Euclidean distance between vectors and matrices.

This article displays some simple examples below. For calling the RNN and examining the classifier's performance, the following packages are required: **keras** (Allaire and Chollet 2017), **dummies** (Brown 2012) and **pROC** (Robin, Turck, Hainard, Tiberti, Lisacek, Sanchez, and MÃijller 2011).

## 5. Examples

### 5.1. Data loading oversampling

The **OSTSC** package provides three small built-in datasets for verification that the package has correctly installed and generates balanced time series. The first two examples use `OSTSC()` to balance binary data while the third balances multinomial data.

*The synthetically generated control dataset*

The dataset Dataset_Synthetic_Control is a time series of sensor measurements of human body motion generated by Alcock, Manolopoulos, Laboratory, and Informatics (1999). We introduce the following labeling: Class 1 represents the 'Normal' state, while Class 0 represents one of 'Cyclic', 'Increasing trend', 'Decreasing trend', 'Upward shift' or 'Downward shift' (Pham and Chan 1998). Users load the dataset by calling `data()`.

```
R> data(Dataset_Synthetic_Control)
R> train.label <- Dataset_Synthetic_Control$train.y
R> train.sample <- Dataset_Synthetic_Control$train.x
R> test.label <- Dataset_Synthetic_Control$test.y
R> test.sample <- Dataset_Synthetic_Control$test.x
```

Each row of the dataset is a sequence of observations. The sequence is of length 60 and there are 300 observations.

```
R> dim(train.sample)
```

```
[1] 300  60
```

The imbalance ratio of the training data is 1:5.

```
R> table(train.label)
```

```
train.label
  0   1
250  50
```

We now provide a simple example demonstrating oversampling of the minority data to match the number of observations of the majority class. The output 'MyData' stores the samples (a.k.a. features) and labels. There are ten parameters passed to `OSTSC()`, the details of which can be found in the help documentation. Calling `OSTSC()` requires the user to provide at least the labels and sample data - the other parameters have default values. It is important to note that the labels are separated from the samples.

```
R> MyData <- OSTSC(train.sample, train.label, parallel = FALSE)
R> over.sample <- MyData$sample
R> over.label <- MyData$label
```

The positive and negative observations are now balanced. Let us check the (im)balance of the new dataset.

```
R> table(over.label)
```

```
over.label
  0   1
250 250
```

The minority class data is oversampled to produce a balanced feature set. The minority-majority formation uses a one-vs-rest strategy. For this binary dataset, the Class 1 data has been oversampled to yield the same number of observations as Class 0.

```
R> dim(over.sample)
```

```
[1] 500  60
```

### *The automatic diatoms identification dataset*

The dataset Dataset_Adiac is generated from a pilot study identifying diatoms (unicellular algae) from images by Jalba, Wilkinson, and Roerdink (2004) originally has 37 classes. For the purpose of demonstrating **OSTSC** we selected only one class as the positive class (Class 1) and all others are set as the negative class (Class 0) to form a highly imbalanced dataset. Users load the dataset into R by calling `data()`.

```
R> data(Dataset_Adiac)
R> train.label <- Dataset_Adiac$train.y
R> train.sample <- Dataset_Adiac$train.x
R> test.label <- Dataset_Adiac$test.y
R> test.sample <- Dataset_Adiac$test.x
```

The training dataset consists of 390 observations of a 176 length sequence.

```
R> dim(train.sample)

[1] 390 176
```

The imbalance ratio of the training data is 1:29.

```
R> table(train.label)

train.label
  0   1
377  13
```

The `OSTSC()` generates a balanced dataset:

```
R> MyData <- OSTSC(train.sample, train.label, parallel = FALSE)
R> over.sample <- MyData$sample
R> over.label <- MyData$label
```

`table()` provides a summary of the balanced dataset.

```
R> table(over.label)

over.label
  0   1
377 377
```

### *The high frequency trading dataset*

`OSTSC()` provides support for multinomial classification. The user specifies which classes should be oversampled. Typically, oversampling is first applied to the minority class - the class with the least number of observations. The dataset Dataset_HFT300 is extracted from a real high frequency trading datafeed (Dixon 2017). It contains a feature representing instantaneous liquidity imbalance using the best bid to ask ratio. The data is labeled so that $Y = 1$ for a next event mid-price up-tick, $Y = -1$ for a down-tick, and $Y = 0$ for no mid-price movement.Users load the dataset into the R environment by calling `data()`.

```
R> data(Dataset_HFT300)
R> train.label <- Dataset_HFT300$y
R> train.sample <- Dataset_HFT300$x
```

The sequence length is set to 10 and 300 sequence observations are randomly drawn for this example dataset.

```
R> dim(train.sample)

[1] 300  10
```

The imbalance ratio of the three class dataset is 1:48:1.

```
R> table(train.label)

train.label
 -1   0   1
  6 288   6
```

This example demonstrates the case when there are two minority classes and both are oversampled. The oversampling is processed using a one-vs-rest strategy, which means that each minority class is oversampled to the same count as the sum of the count of all other classes. This results in a slight imbalance in the total number of labels.

```
R> MyData <- OSTSC(train.sample, train.label, parallel = FALSE)
R> over.sample <- MyData$sample
R> over.label <- MyData$label
```

We observe the ratio of the classes after oversampling.

```
R> table(over.label)

over.label
 -1   0   1
294 288 294
```

The above examples illustrate how `OSTSC()` oversamples small datasets. In the next section, we demonstrate and evaluate the oversampled data on two medium size datasets.

### 5.2. Applying OSTSC to medium size datasets

*The Electrical Devices dataset*

The dataset Dataset_ElectricalDevices is a sample collected from the 'Powering the Nation' study (Lines, Bagnall, Caiger-Smith, and Anderson 2011). This study seeks to reduce the UK's carbon footprint by collecting behavioural data on how consumers use electricity within the home. Each class represent a signal from a different electrical device. Classes 5 and 6 in the original dataset are set as the negative and positive respectively. The dataset is split into training and testing feature vectors and labels.

```
R> ElectricalDevices <- Dataset_ElectricalDevices()
R> train.label <- ElectricalDevices$train.y
R> train.sample <- ElectricalDevices$train.x
R> test.label <- ElectricalDevices$test.y
R> test.sample <- ElectricalDevices$test.x
R> vali.label <- ElectricalDevices$vali.y
R> vali.sample <- ElectricalDevices$vali.x
```

Each row in the data represents a sequence of length 96.

```
R> dim(train.sample)

[1] 2200   96
```

The imbalance ratio of the training data is 1:21.

```
R> table(train.label)

train.label
   0    1
2100  100
```

After oversampling with OSTSC, the positive and negative observations are balanced.

```
R> MyData <- OSTSC(train.sample, train.label, parallel = FALSE)
R> over.sample <- MyData$sample
R> over.label <- MyData$label

R> table(over.label)

over.label
   0    1
2100 2100
```

An LSTM classifier is used as the basis for performance assessment of oversampling with OSTSC. We use the **keras** package (Allaire and Chollet 2017) to configure the architecture, hyper-parameters and learning schedule of the LSTM classifier for sequence classification.

As a baseline for **OSTSC**, we assess the performance of LSTM trained on the unbalanced and balanced data. All performances are evaluated out-of-sample unless stated otherwise. Note that for the multi-classification examples, each F1 history is shown separately but is evaluated on the same validation set during training. The procedure for applying Keras is next outlined:

One-hot encode the categorical label vectors as binary class matrices using `dummy()`. Then transform the feature matrices to tensors for LSTM: Initialize a sequential model, add layers and then compile it. Train the LSTM classifier on both of the imbalanced and the oversampled data.

```
R> model <- keras_model_sequential()
R> model %>%
+      layer_lstm(10, input_shape = c(dim(train.x)[2], dim(train.x)[3])) %>%
+      #layer_dropout(rate = 0.1) %>%
+      layer_dense(dim(train.y)[2]) %>%
+      layer_dropout(rate = 0.1) %>%
+      layer_activation("softmax")
R> history <- LossHistory$new()
R> model %>% compile(
+      loss = "categorical_crossentropy",
+      optimizer = optimizer_adam(lr = 0.005),
+      metrics = c("accuracy",'f1_score_0' = metric_f1_0, 'f1_score_1' = metric_f1_1)
+  )
R> lstm.before <- model %>% fit(
+      x = train.x,
+      y = train.y,
+      validation_data=list(vali.x,vali.y),
+      batch_size = 256,
+      callbacks = list(history),
+      epochs = 50
+  )

R> model.over <- keras_model_sequential()
R> model.over %>%
+      layer_lstm(10, input_shape = c(dim(over.x)[2], dim(over.x)[3])) %>%
+      #layer_dropout(rate = 0.1) %>%
+      layer_dense(dim(over.y)[2]) %>%
+      layer_dropout(rate = 0.1) %>%
+      layer_activation("softmax")
R> history.over <- LossHistory$new()
R> model.over %>% compile(
+      loss = "categorical_crossentropy",
+      optimizer = optimizer_adam(lr = 0.005),
+      metrics = c("accuracy",'f1_score_0' = metric_f1_0, 'f1_score_1' = metric_f1_1)
+  )
R> lstm.after <- model.over %>% fit(
+      x = over.x,
+      y = over.y,
+      validation_data=list(vali.x,vali.y),
+      batch_size = 256,
+      callbacks = list(history.over),
+      epochs = 50
+  )
```

From the training history, Figures 1 and 2 compare the F1 scores of the two models without and with oversampling. Figure 3 compares the losses of the two models.

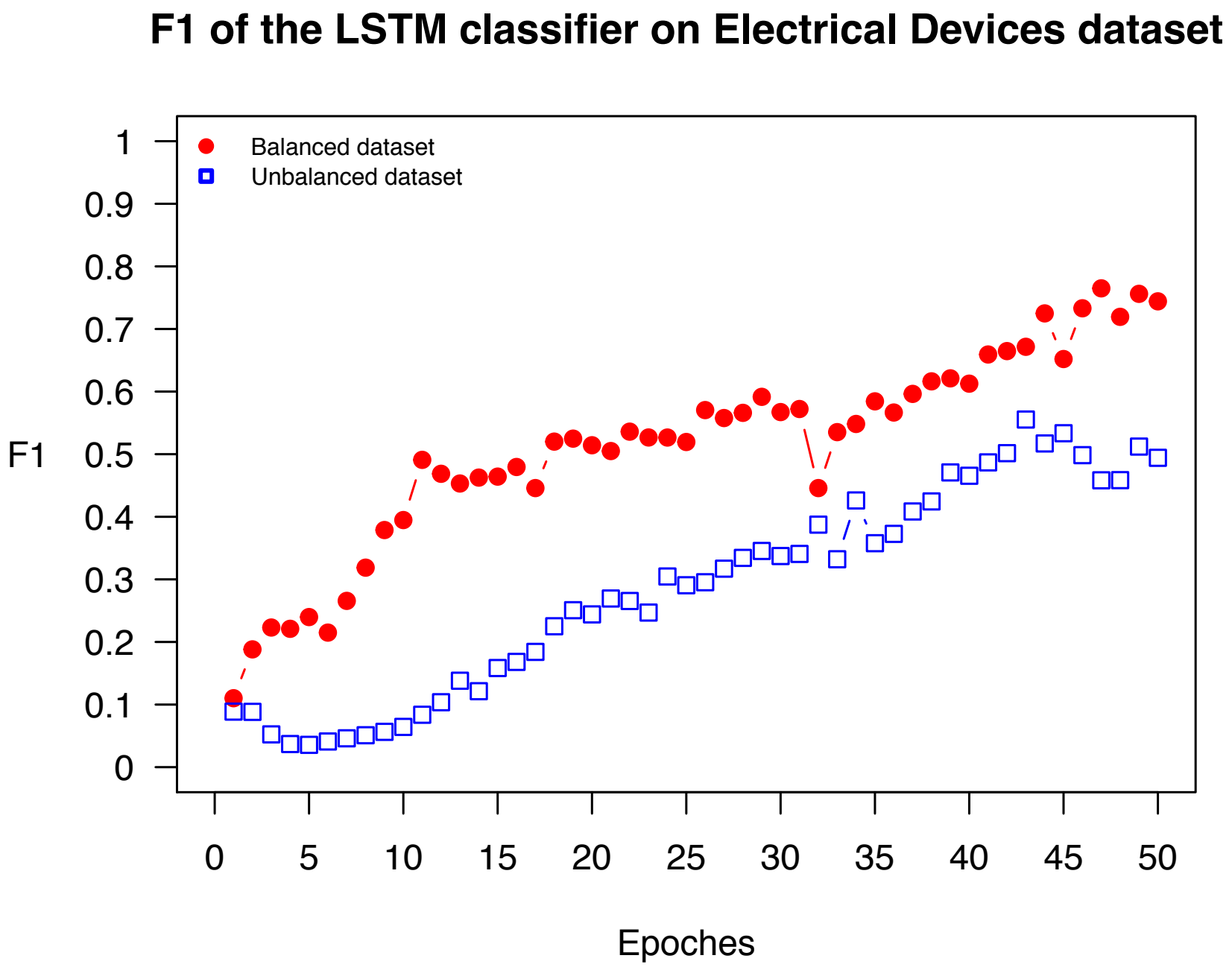

Figure 1: The F1 scores (class 1) of the LSTM classifier trained on the unbalanced and balanced Electrical Devices dataset. Both metrics are evaluated at the end of each epoch.

In addition to the training history, Figures 4 and 5 compare the confusion matrices of the two models without and with oversampling. Figure 6 compares the receiver operating characteristic (ROC) curves of the models. The final out-of-sample F1 scores of the two trained models are also shown below for comparison.

```
The class 1 F1 score without oversampling:  0.6712329

The class 0 F1 score without oversampling:  0.9817768

The class 1 F1 score with oversampling:  0.7368421

The class 0 F1 score with oversampling:  0.9847793
```

*The Electrocardiogram dataset*

The dataset *Dataset_ECG* was originally created by (Goldberger, Amaral, Glass, Hausdorff, Ivanov, Mark, Mietus, Moody, Peng, and Stanley 2000) and records heartbeats from patients with severe congestive heart failure. The dataset was pre-processed to extract heartbeat sequences and add labels by (Chen, Hao, Rakthanmanon, Zakaria, Hu, and Keogh 2015). The article uses 5,000 randomly selected heartbeat sequences.

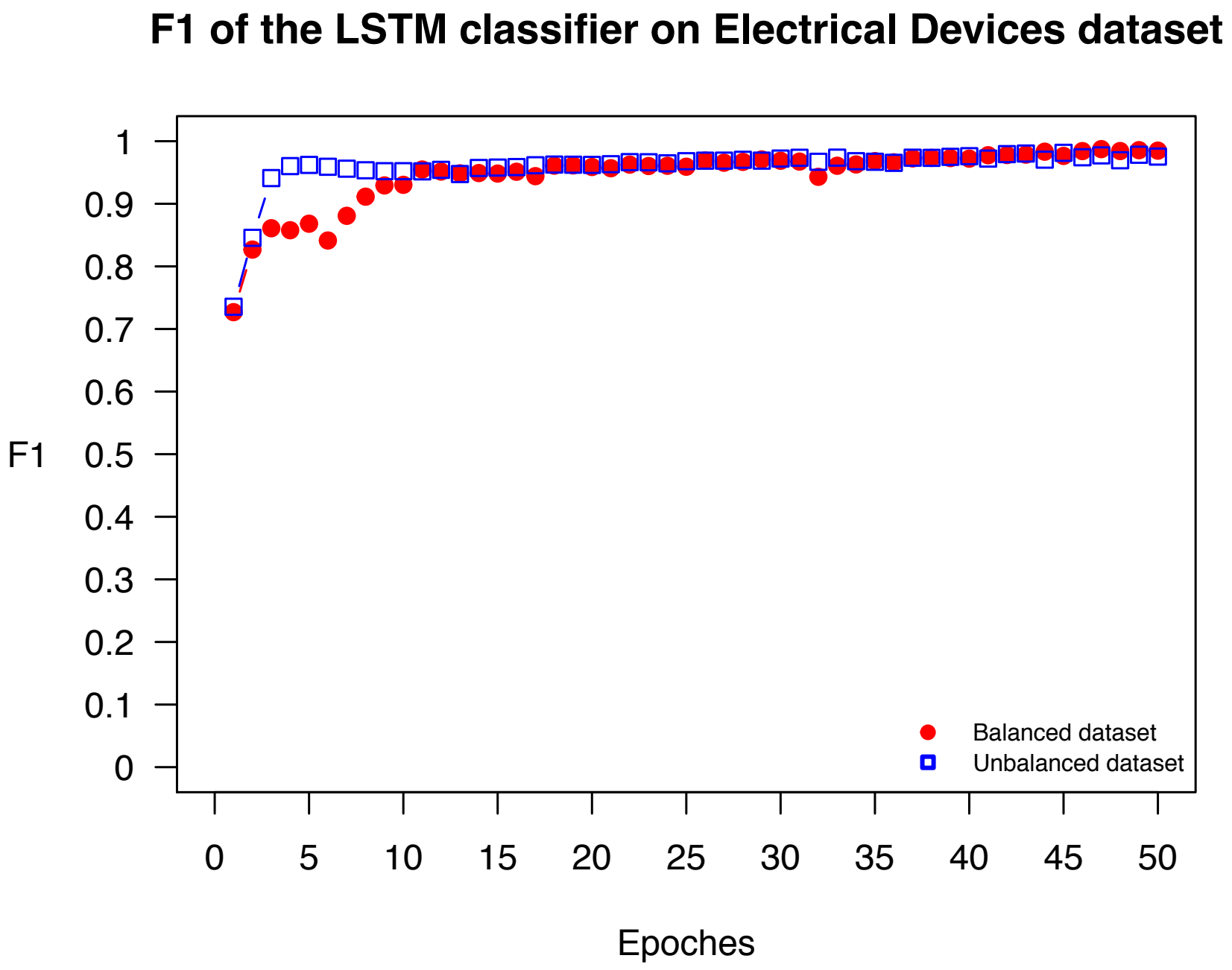

Figure 2: The F1 scores (class 0) of the LSTM classifier trained on the unbalanced and balanced Electrical Devices dataset. Both metrics are evaluated at the end of each epoch.

```
R> ECG <- Dataset_ECG()
R> train.label <- ECG$train.y
R> train.sample <- ECG$train.x
R> test.label <- ECG$test.y
R> test.sample <- ECG$test.x
R> vali.label <- ECG$vali.y
R> vali.sample <- ECG$vali.x
```

Each row in the data represents a sequence of length 140.

```
R> dim(train.sample)
```

```
[1] 2296  140
```

This experiment uses 3 classes of the dataset to ensure a high degree of imbalance: the imbalance ratio is 32:1:2.

```
R> table(train.label)
```

```
train.label
   0    1    2 
2100   66  130 
```

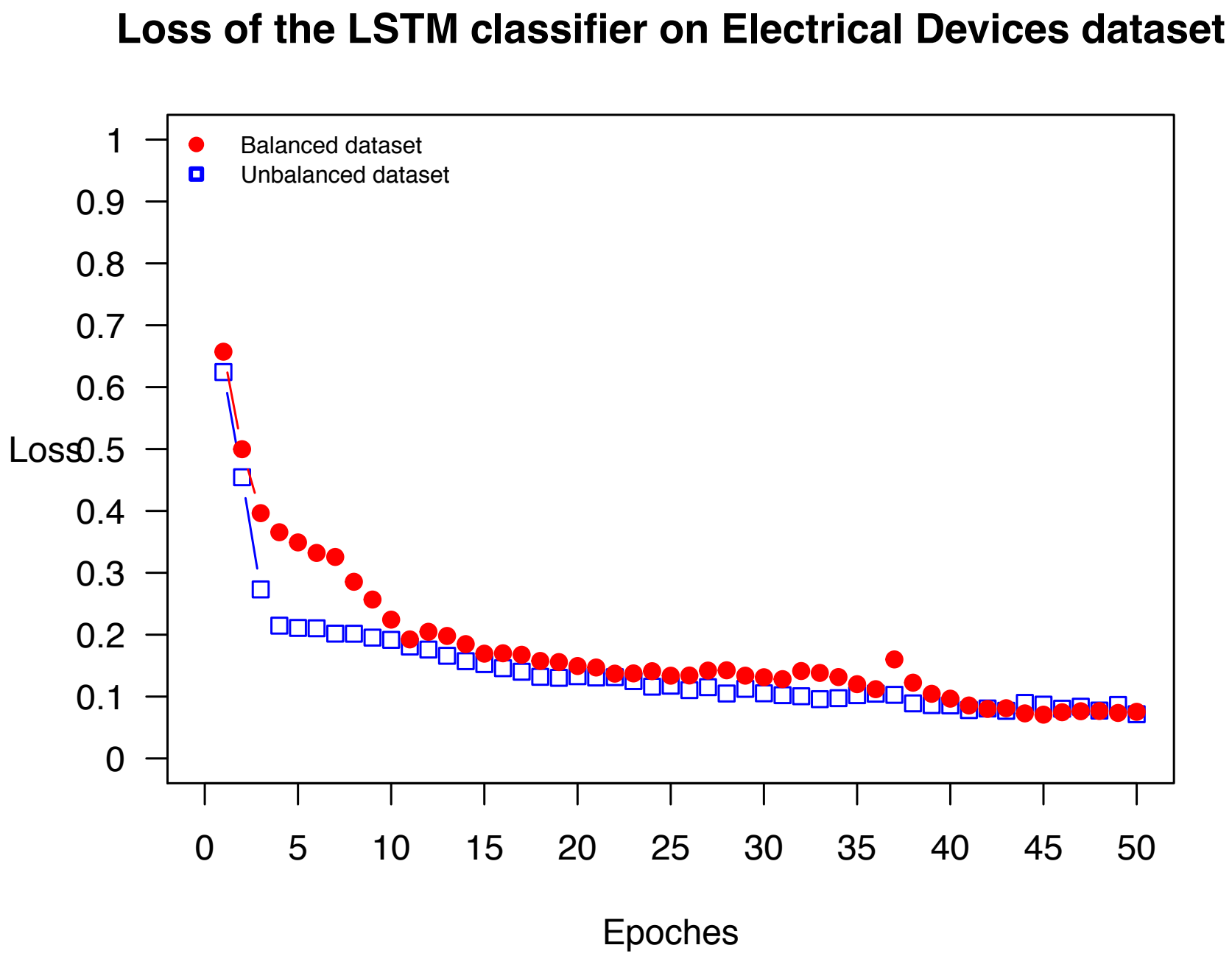

Figure 3: The losses of the LSTM classifier trained on the unbalanced and balanced Electrical Devices dataset. Both metrics are evaluated at the end of each epoch.

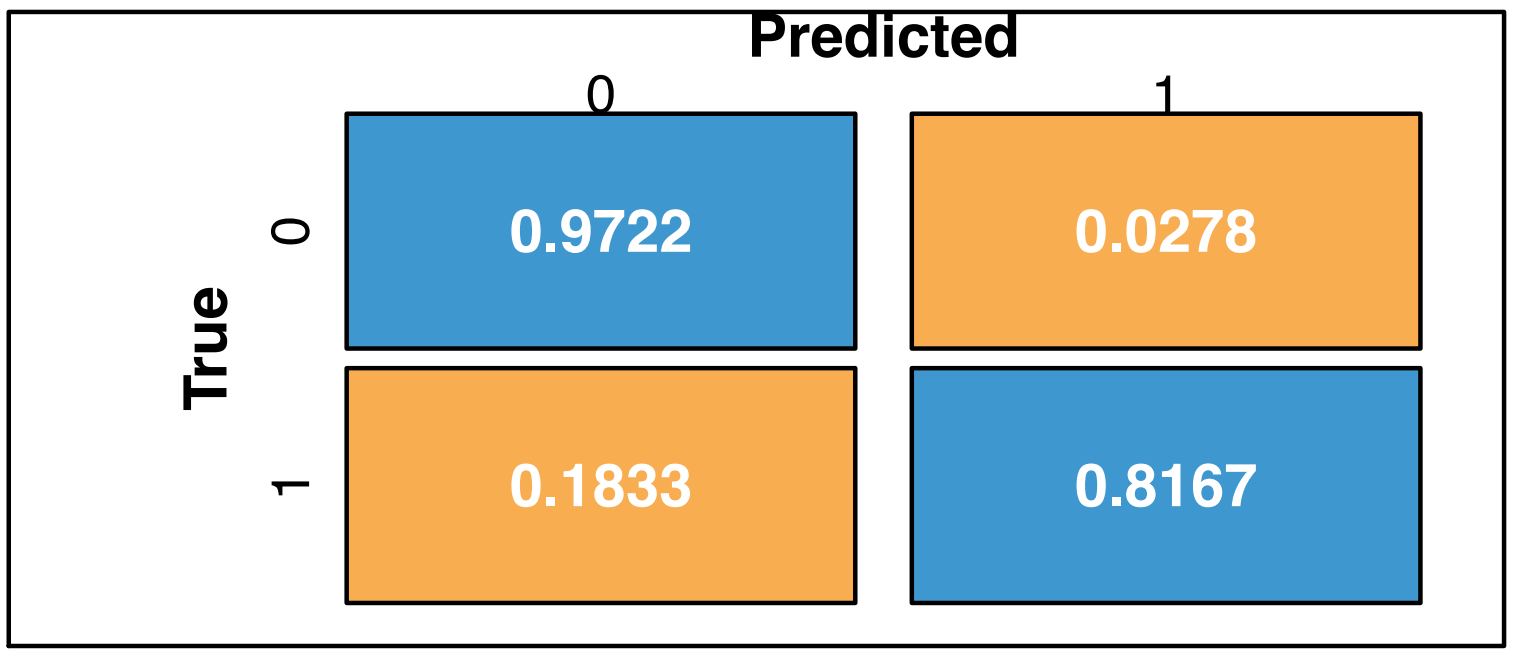

Figure 4: Normalized confusion matrix of LSTM applied to the Electrical Devices dataset without oversampling.

Let us check that the data is balanced after oversampling.

```
R> MyData <- OSTSC(train.sample, train.label, parallel = FALSE)
R> over.sample <- MyData$sample
```

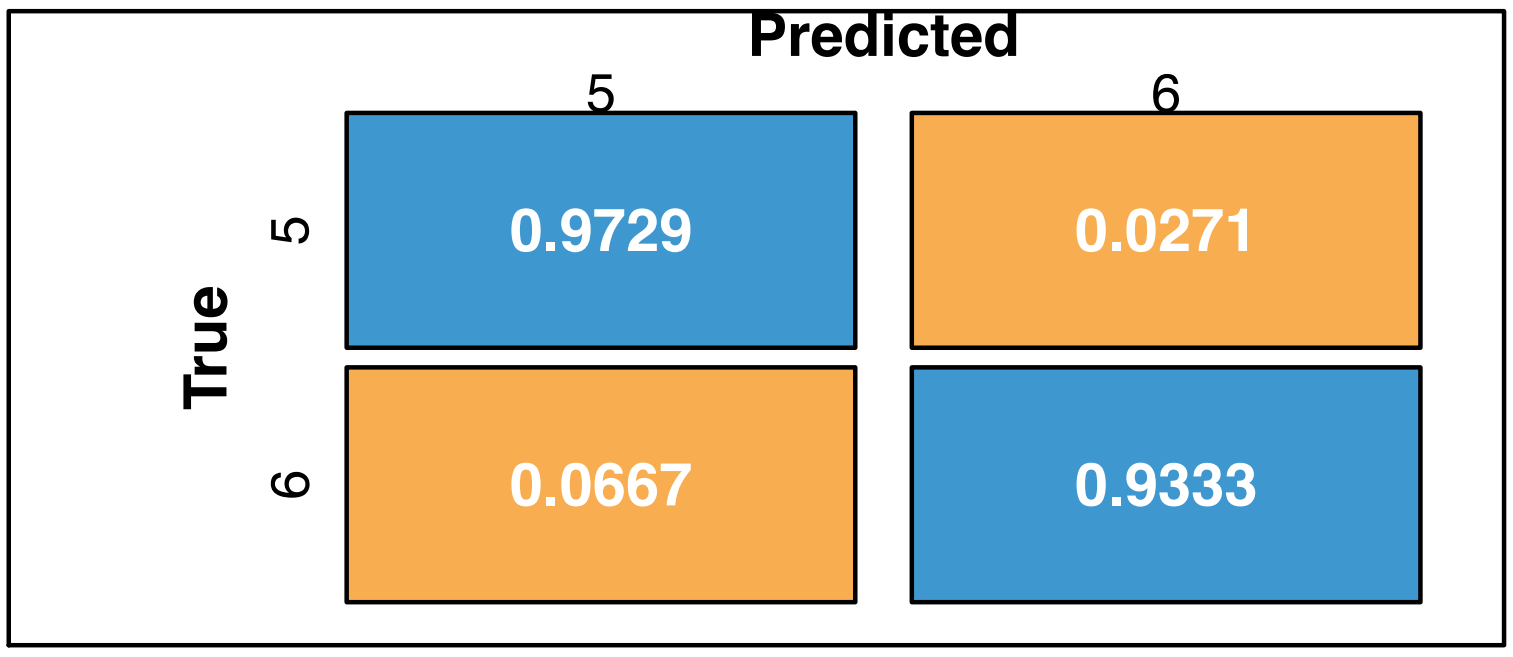

Figure 5: Normalized confusion matrix of LSTM applied to the Electrical Devices dataset with oversampling.

```
R> over.label <- MyData$label

R> table(over.label)

over.label
   0    1    2
2100 2230 2166
```

We evaluate the effect of oversampling on the performance of LSTM following Steps 1-3 above. First the data is transformed. During configuring and training the model, the F1 scores and losses are measured at the end of each epoch using the same validation set.

```
R> model <- keras_model_sequential()
R> model %>%
+      layer_lstm(10, input_shape = c(dim(train.x)[2], dim(train.x)[3])) %>%
+      #layer_dropout(rate = 0.1) %>%
+      layer_dense(dim(train.y)[2]) %>%
+      layer_dropout(rate = 0.1) %>%
+      layer_activation("softmax")
R> history <- LossHistory$new()
R> model %>% compile(
+      loss = "categorical_crossentropy",
+      optimizer = optimizer_adam(lr = 0.001),
+      metrics = c("accuracy",'f1_score_0' = metric_f1_0, 'f1_score_1' = metric_f1_1,
+                  'f1_score_2' = metric_f1_2)
+   )
R> lstm.before <- model %>% fit(
+      x = train.x,
+      y = train.y,
```

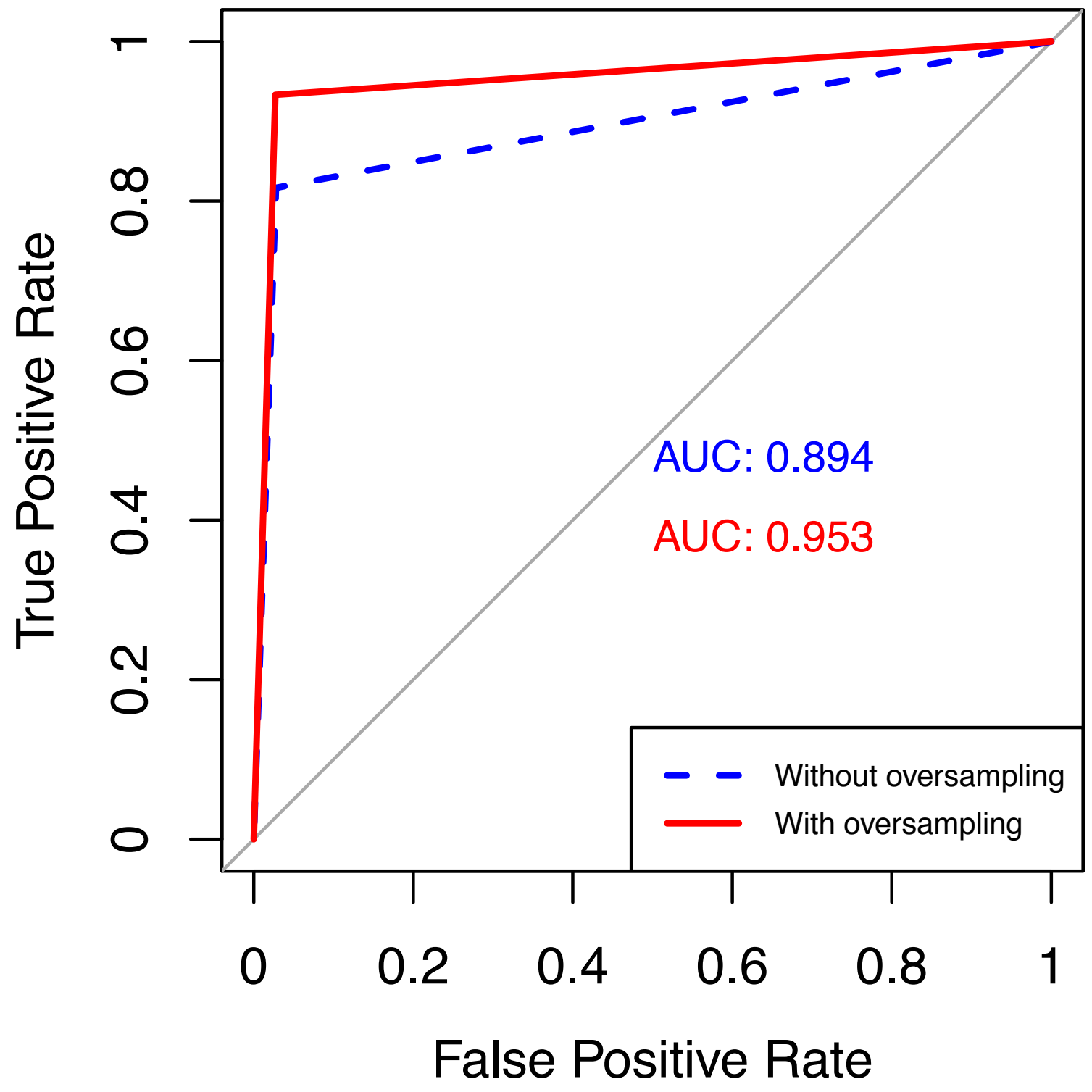

Figure 6: ROC curves comparing the effect of oversampling on the performance of LSTM applied to the Electrical Devices dataset.

```
+       validation_data=list(vali.x,vali.y),
+       batch_size = 256,
+       callbacks = list(history),
+       epochs = 50
+   )

R> model.over <- keras_model_sequential()
R> model.over %>%
+       layer_lstm(10, input_shape = c(dim(over.x)[2], dim(over.x)[3])) %>%
+       #layer_dropout(rate = 0.1) %>%
+       layer_dense(dim(over.y)[2]) %>%
+       layer_dropout(rate = 0.1) %>%
+       layer_activation("softmax")
R> history.over <- LossHistory$new()
R> model.over %>% compile(
+       loss = "categorical_crossentropy",
```

```
+       optimizer = optimizer_adam(lr = 0.001),
+       metrics = c("accuracy",'f1_score_0' = metric_f1_0, 'f1_score_1' = metric_f1_1,
+                   'f1_score_2' = metric_f1_2)
+   )
R> lstm.after <- model.over %>% fit(
+       x = over.x,
+       y = over.y,
+       validation_data=list(vali.x,vali.y),
+       batch_size = 256,
+       callbacks = list(history.over),
+       epochs = 50
+   )
```

Keeping the number of epoches fixed, Figures 7, 8 and 9 respectively compare the F1 scores of three different classes of the two models without and with oversampling. Figure 10 compares the losses of the two models. From the losses and F1 scores, we note that the model has not yet been adequately trained after 50 epoches. We are trying to demonstrate the utility of OSTSC with only a modest amount of computation. The user can of course choose to increase the number of epoches, but will this require more computation. The user should refer to the larger dataset examples below for comparative evaluations which use more epoches for training LSTM.

In addition to the training history, Figures 11 and 12 compare the confusion matrices of the two models without and with oversampling. Figure 13 compares the receiver operating characteristic (ROC) curves of the models. The final F1 scores of the two trained models, using the same validation set, are also shown below for comparison.

```
The class 2 F1 score without oversampling:  0.36

The class 1 F1 score without oversampling:  0.64

The class 0 F1 score without oversampling:  0.9698858

The class 2 F1 score with oversampling:  0.6969697

The class 1 F1 score with oversampling:  0.5909091

The class 0 F1 score with oversampling:  0.9784483
```

### 5.3. Evaluating OSTSC on the large datasets

The evaluation of oversampling uses larger datasets: the MHEALTH and HFT datasets. The purpose of this evaluation is to demonstrate how **OSTSC** performs at scale. We increase the data sizes by a factor of up to 10x. The evaluation of each dataset takes approximately three hours on a 1.7 GHz four-core laptop with 8GM of RAM.

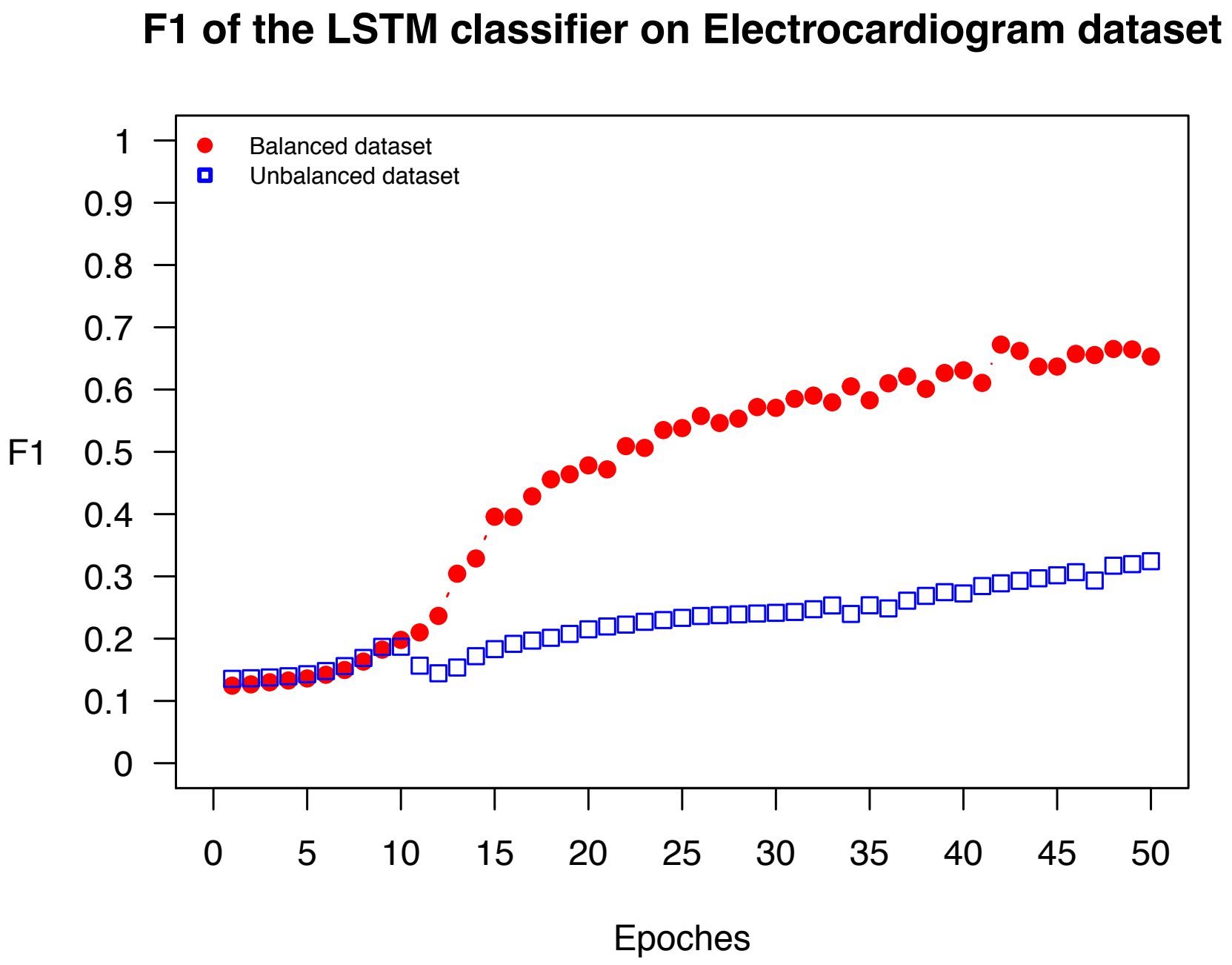

Figure 7: The F1 scores (class 2) of the LSTM classifier trained on the unbalanced and balanced Electrocardiogram dataset. Both metrics are evaluated at the end of each epoch.

*The MHEALTH dataset*

The dataset Dataset_MHEALTH benchmarks techniques for human behavioral analysis applied to multimodal body sensing (Banos, Garcia, Holgado-Terriza, Damas, Pomares, Rojas, Saez, and Villalonga 2014). In this experiment, only Subjects 1-5 and Feature 12 (the x coordinate of the magnetometer reading from the left-ankle sensor) are used. The dataset is labeled with a dichotonomous response Banos2015. Class 11 (Running) is set as the positive and the remaining states are the negative. The dataset is split into training and testing feature vectors and labels.

```
R> MHEALTH <- Dataset_MHEALTH()
R> train.label <- MHEALTH$train.y
R> train.sample <- MHEALTH$train.x
R> test.label <- MHEALTH$test.y
R> test.sample <- MHEALTH$test.x
R> vali.label <- MHEALTH$vali.y
R> vali.sample <- MHEALTH$vali.x
```

Each row in the data represents a sequence of length 30.

```
R> dim(train.sample)
```

```
[1] 10250    30
```

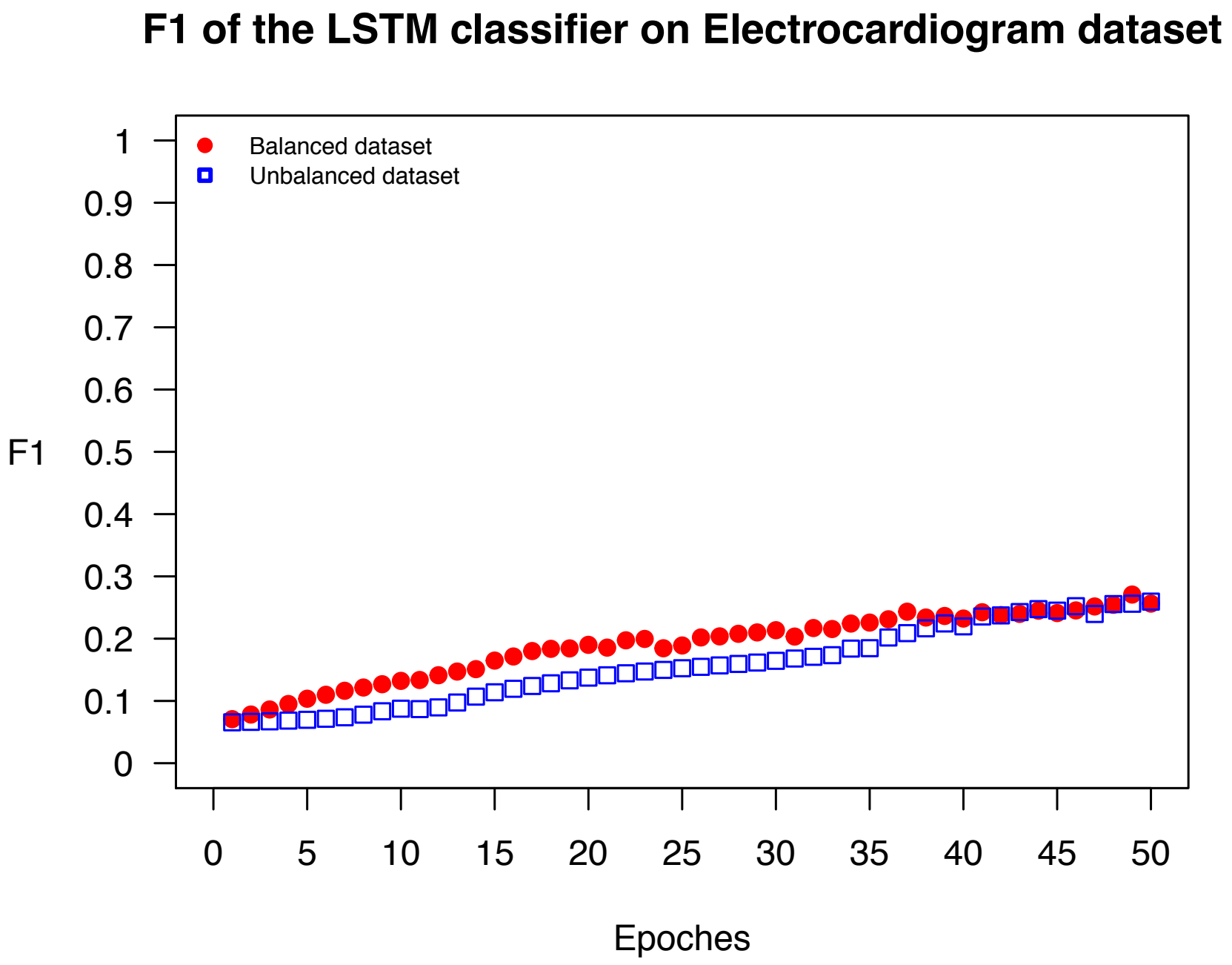

Figure 8: The F1 scores (class 1) of the LSTM classifier trained on the unbalanced and balanced Electrocardiogram dataset. Both metrics are evaluated at the end of each epoch.

Class 1 represents the positive data and class 0 represents the negative. The imbalance ratio of the train dataset is 1:40.

```
R> table(train.label)
```

```
train.label
    0     1
10000   250
```

After oversampling, the positive and negative observations are balanced.

```
R> MyData <- OSTSC(train.sample, train.label, parallel = FALSE)
R> over.sample <- MyData$sample
R> over.label <- MyData$label
```

```
R> table(over.label)
```

```
over.label
    0     1
10000 10000
```

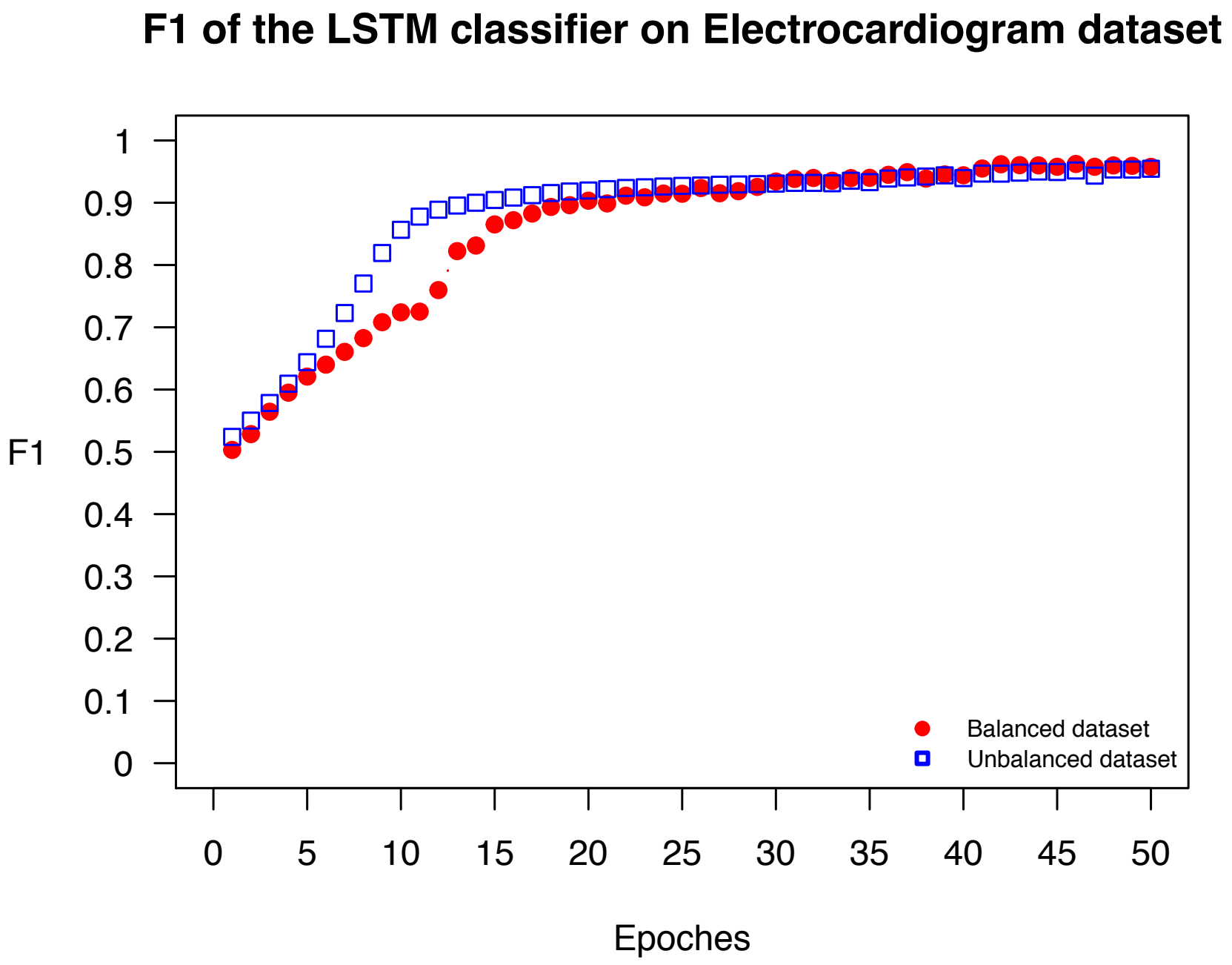

Figure 9: The F1 scores (class 0) of the LSTM classifier trained on the unbalanced and balanced Electrocardiogram dataset. Both metrics are evaluated at the end of each epoch.

We are concerned more here with the comparative performance without and with oversampling and less with the absolute gain (which is subject to further parameter tuning). Keeping the number of epoches fixed, Figures 14 and 15 compare the F1 scores of the two models without and with oversampling, Figure 16 compares the losses of the two models, Figures 17 and 18 compare the confusion matrices of the two models without and with oversampling, and Figure 19 compares the ROC curves of the models. The final F1 scores of the two trained models, using the same validation set, are also shown below for comparison.

```
R> train.y <- dummy(train.label)
R> test.y <- dummy(test.label)
R> train.x <- array(train.sample, dim = c(dim(train.sample),1))
R> test.x <- array(test.sample, dim = c(dim(test.sample),1))
R> vali.y <- dummy(vali.label)
R> vali.x <- array(vali.sample, dim = c(dim(vali.sample),1))
R> over.y <- dummy(over.label)
R> over.x <- array(over.sample, dim = c(dim(over.sample),1))

R> model <- keras_model_sequential()
R> model %>%
+    layer_lstm(10, input_shape = c(dim(train.x)[2], dim(train.x)[3])) %>%
+    layer_dropout(rate = 0.2) %>%
```

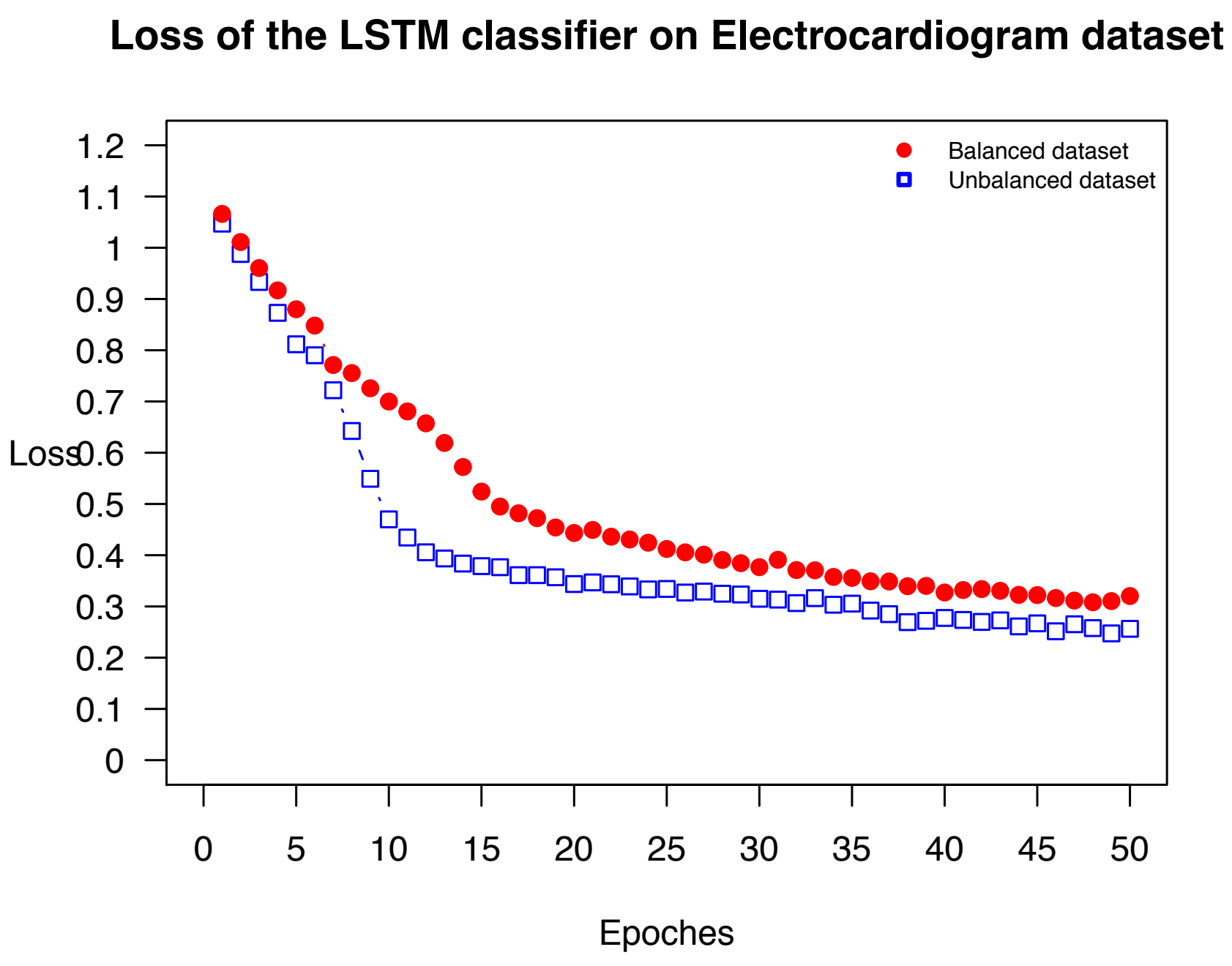

Figure 10: The losses of the LSTM classifier trained on the unbalanced and balanced Electrocardiogram dataset. Both metrics are evaluated at the end of each epoch.

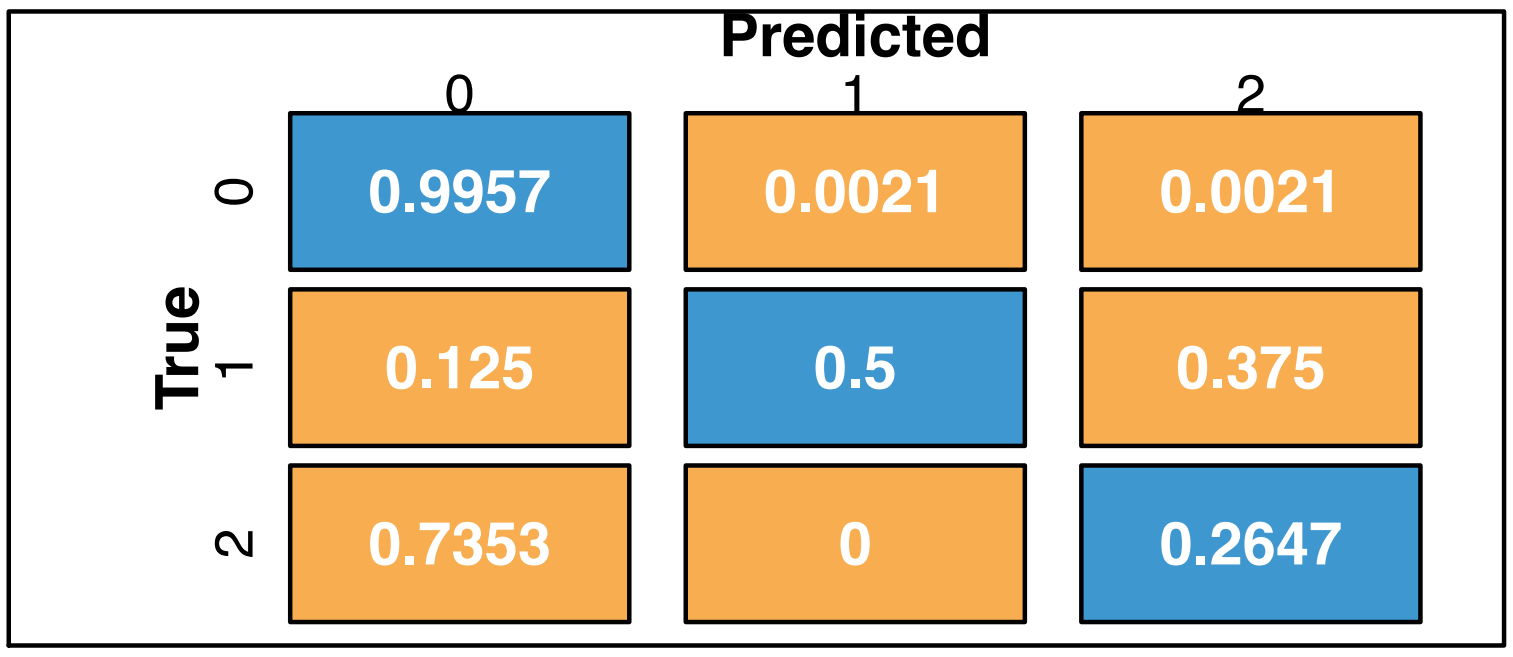

Figure 11: Normalized confusion matrices of LSTM applied to the Electrocardiogram dataset without oversampling.

```
+     layer_dense(dim(train.y)[2]) %>%
+     layer_dropout(rate = 0.2) %>%
+     layer_activation("softmax")
R> history <- LossHistory$new()
```

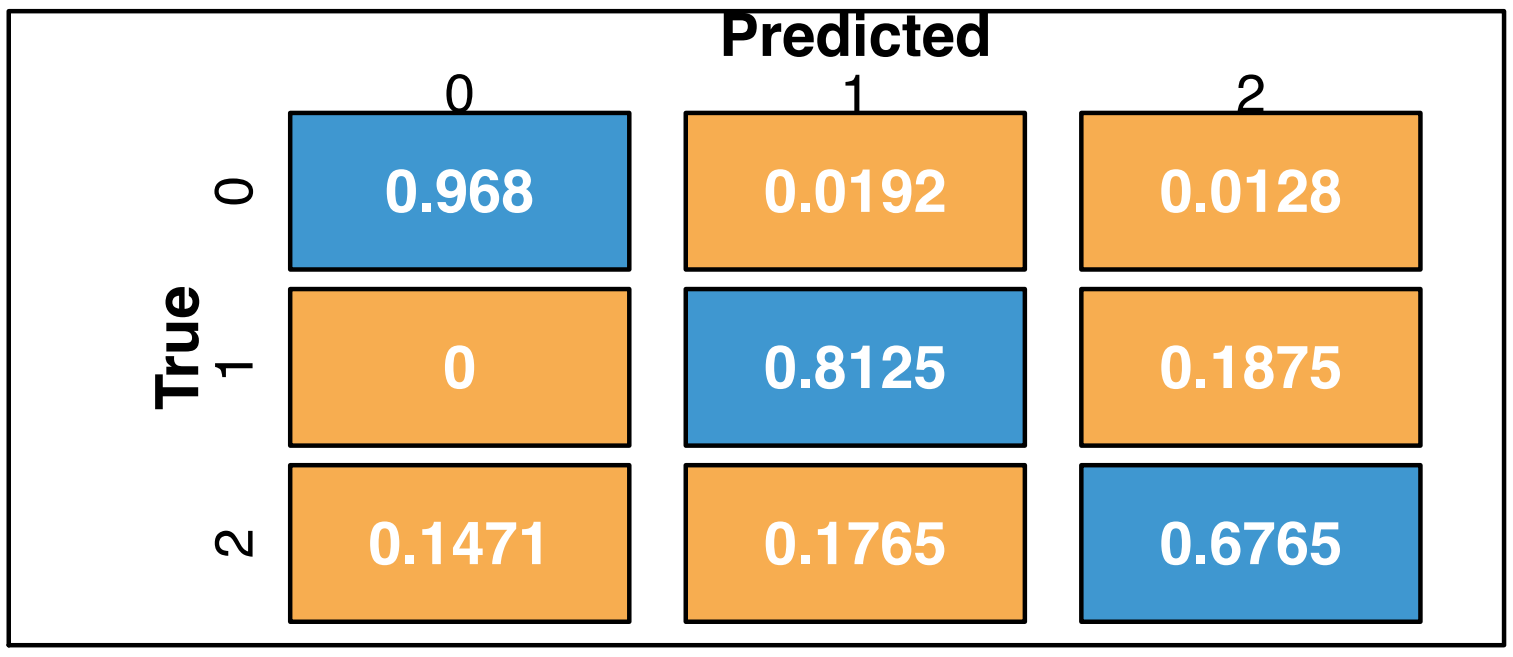

Figure 12: Normalized confusion matrix of LSTM applied to the Electrocardiogram dataset with oversampling.

```
R> model %>% compile(
+    loss = "categorical_crossentropy",
+    optimizer = "adam",
+    metrics = c("accuracy",'f1_score_0' = metric_f1_0, 'f1_score_1' = metric_f1_1)
+  )
R> lstm.before <- model %>% fit(
+    x = train.x,
+    y = train.y,
+    validation_data=list(vali.x,vali.y),
+    callbacks = list(history),
+    epochs = 50
+  )

R> model.over <- keras_model_sequential()
R> model.over %>%
+    layer_lstm(10, input_shape = c(dim(over.x)[2], dim(over.x)[3])) %>%
+    layer_dropout(rate = 0.1) %>%
+    layer_dense(dim(over.y)[2]) %>%
+    layer_dropout(rate = 0.1) %>%
+    layer_activation("softmax")
R> history.over <- LossHistory$new()
R> model.over %>% compile(
+    loss = "categorical_crossentropy",
+    optimizer = "adam",
+    metrics = c("accuracy",'f1_score_0' = metric_f1_0, 'f1_score_1' = metric_f1_1)
+  )
R> lstm.after <- model.over %>% fit(
+    x = over.x,
+    y = over.y,
```

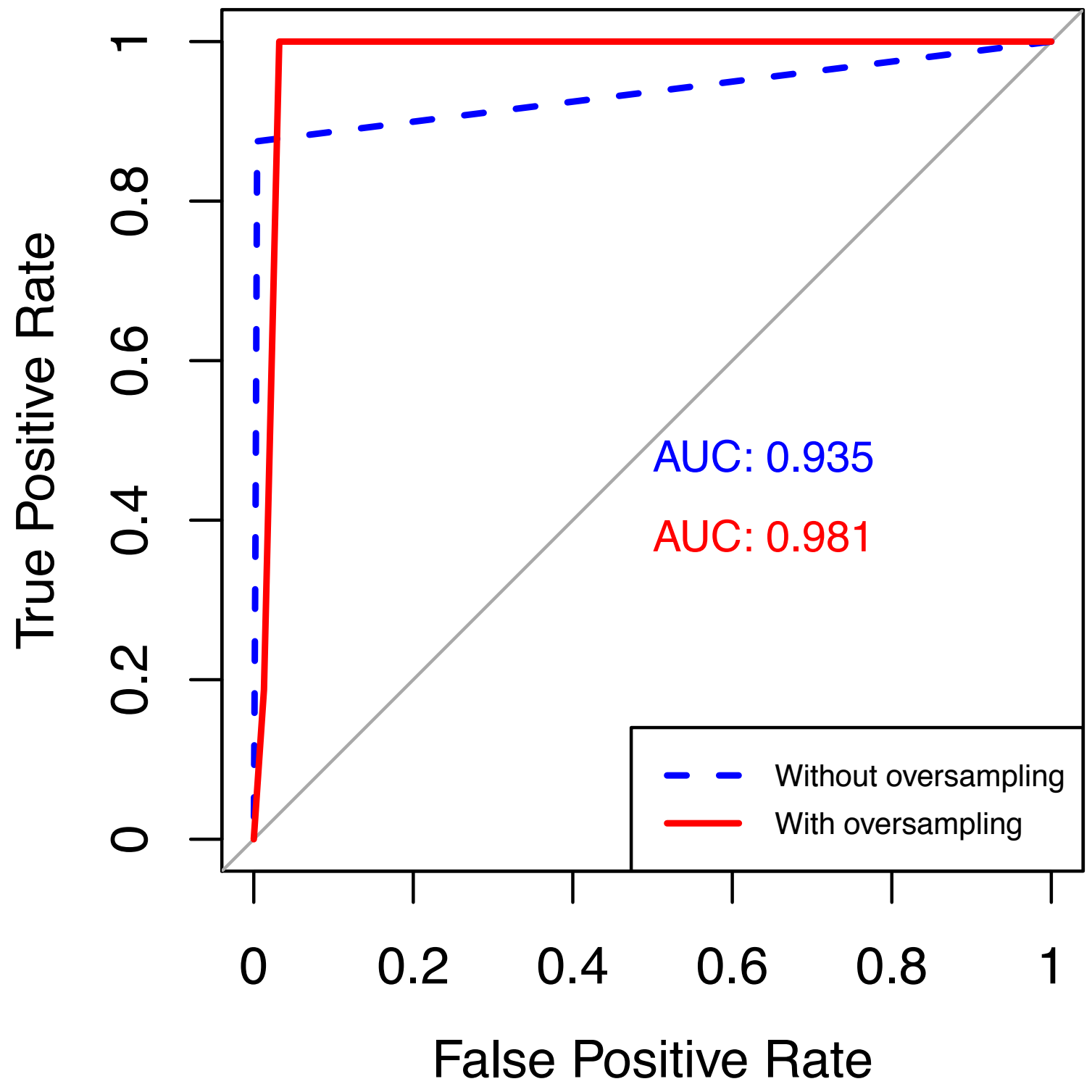

Figure 13: ROC curves of LSTM applied to the Electrocardiogram dataset, with and without oversampling.

```
+     validation_data=list(vali.x,vali.y),
+     callbacks = list(history.over),
+     epochs = 50
+   )
```

```
The class 1 F1 score without oversampling:  0.4496487

The class 0 F1 score without oversampling:  0.985566

The class 1 F1 score with oversampling:  0.493992

The class 0 F1 score with oversampling:  0.9762516
```

*The high frequency trading dataset*

The dataset *Dataset_HFT* has already been introduced in the ‘Data loading  oversampling‘ section. The purpose of this example is to demonstrate the application of oversampling to a large sized dataset consisting of 30,000 observations instead of 300. For control, the imbalance ratio of the dataset is configured to be the same as the smaller dataset. We split the training, validating and testing data by a ratio of 20:3:7.

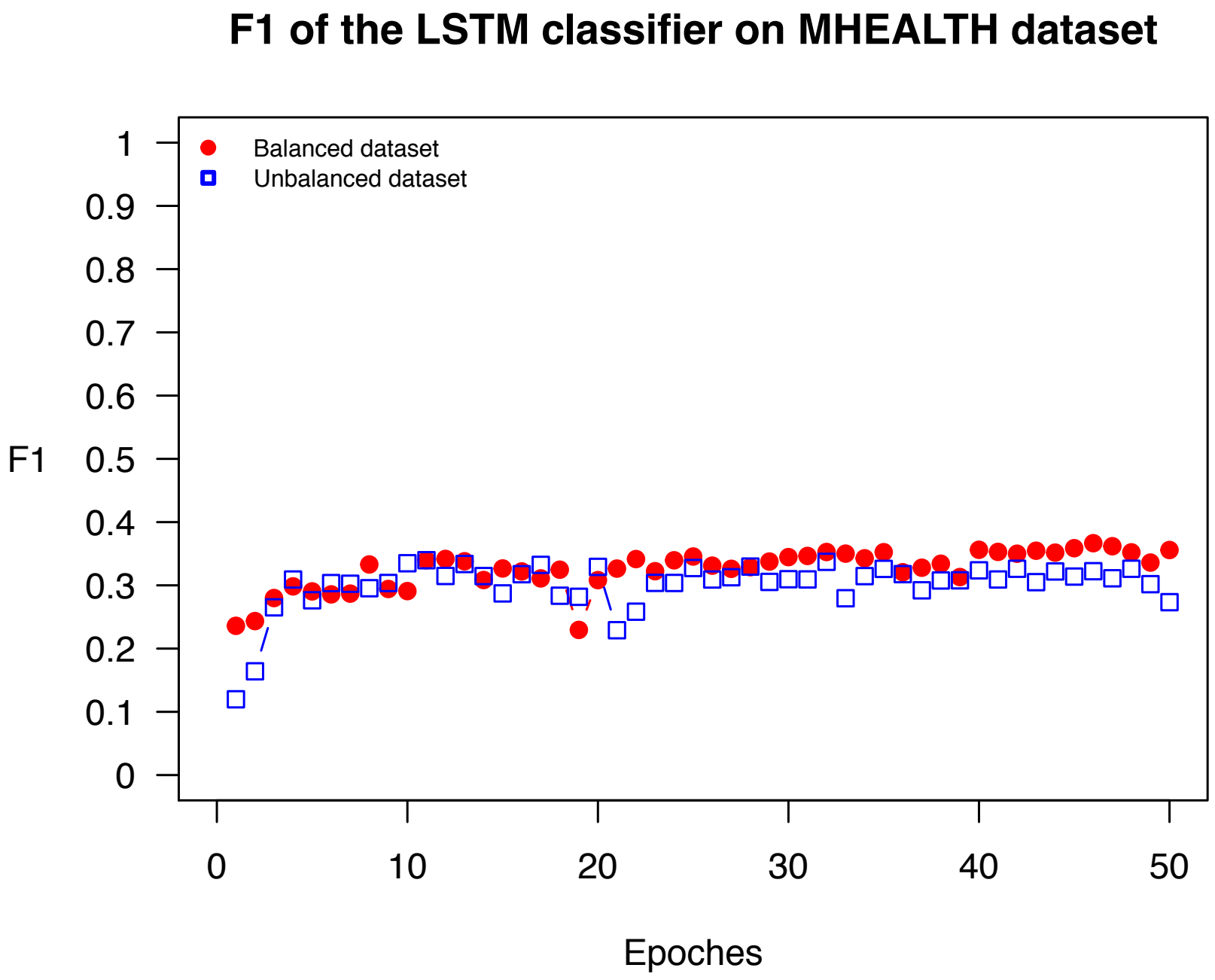

Figure 14: The F1 scores (class 1) of the LSTM classifier trained on the unbalanced and balanced MHEALTH dataset. Both metrics are evaluated at the end of each epoch.

```
R> HFT <- Dataset_HFT()
R> label <- HFT$y
R> sample <- HFT$x
R> train.label <- label[1:20000]
R> train.sample <- sample[1:20000, ]
R> test.label <- label[23001:30000]
R> test.sample <- sample[23001:30000, ]
R> vali.label <- label[20001:23000]
R> vali.sample <- sample[20001:23000, ]
```

The imbalance ratio of the training data is 1:48:1.

```
R> table(train.label)

train.label
   -1     0     1
  383 19269   348
```

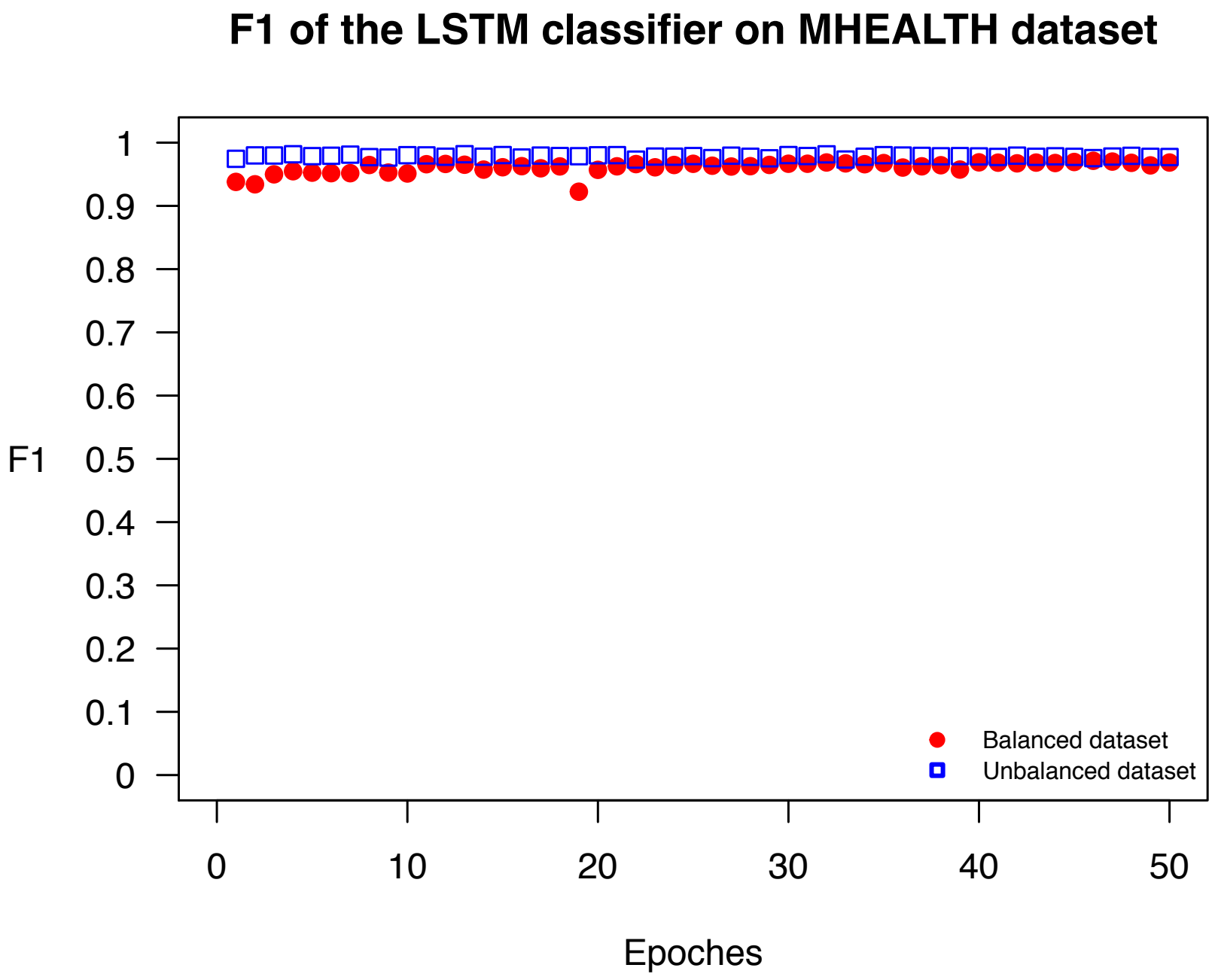

Figure 15: The F1 scores (class 0) of the LSTM classifier trained on the unbalanced and balanced MHEALTH dataset. Both metrics are evaluated at the end of each epoch.

After oversampling the data is balanced.

```
R> MyData <- OSTSC(train.sample, train.label, parallel = FALSE)
R> over.sample <- MyData$sample
R> over.label <- MyData$label

R> table(over.label)
```

```
over.label
   -1     0     1
19617 19269 19652
```

We increase the number of epoches to 100. Figures 20, 21 and 22 compare the F1 scores of the two models without and with oversampling. Figure 23 compares the losses of the two models. Figures 24 and 25 compare the confusion matrices of the two models without and with oversampling. Figure 26 compares the ROC curves of the models. The final F1 scores of the two trained models, using the same validation set, are also shown below for comparison.

```
R> model <- keras_model_sequential()
R> model %>%
+    layer_lstm(10, input_shape = c(dim(train.x)[2], dim(train.x)[3])) %>%
```

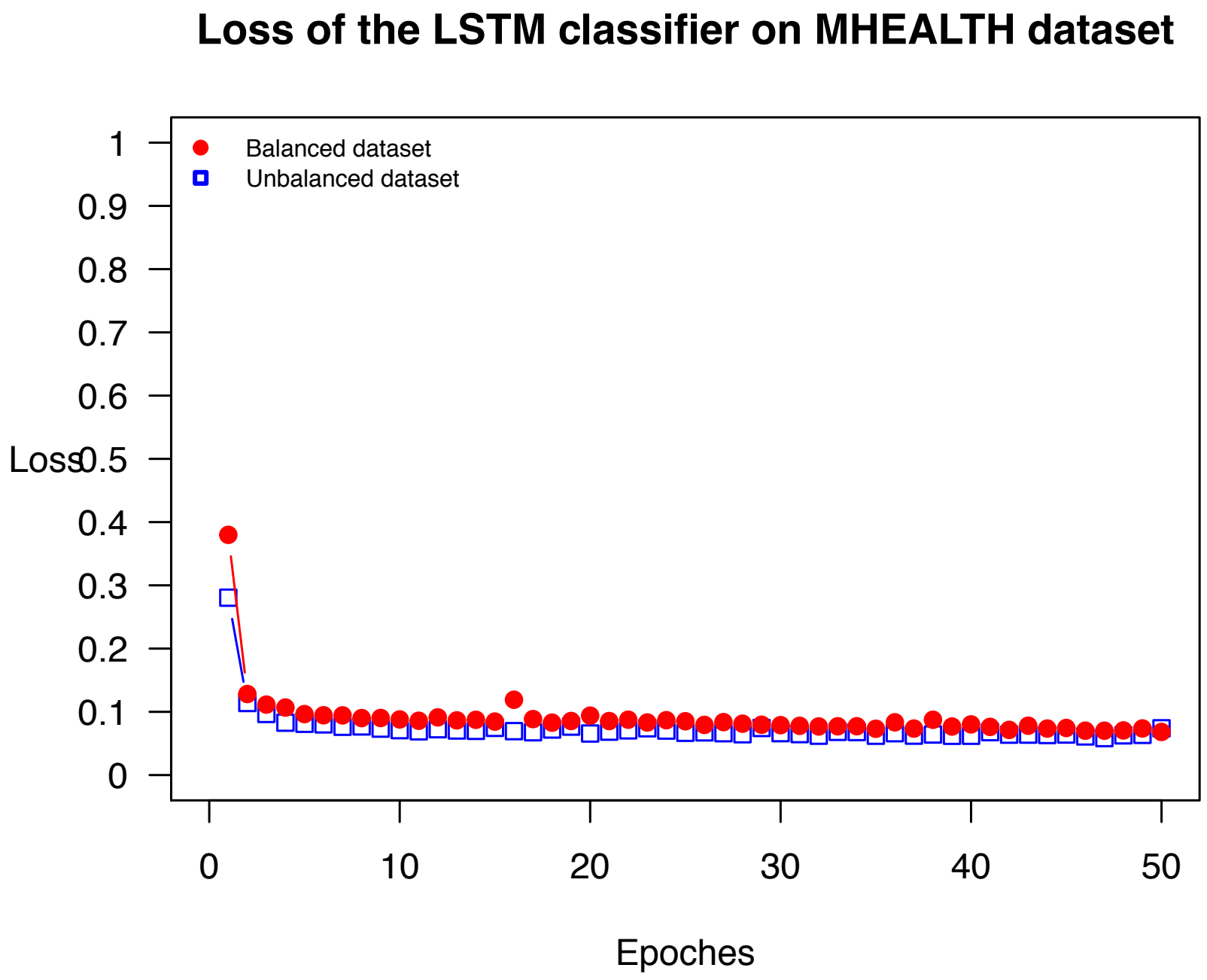

Figure 16: The losses of the LSTM classifier trained on the unbalanced and balanced MHEALTH dataset. Both metrics are evaluated at the end of each epoch.

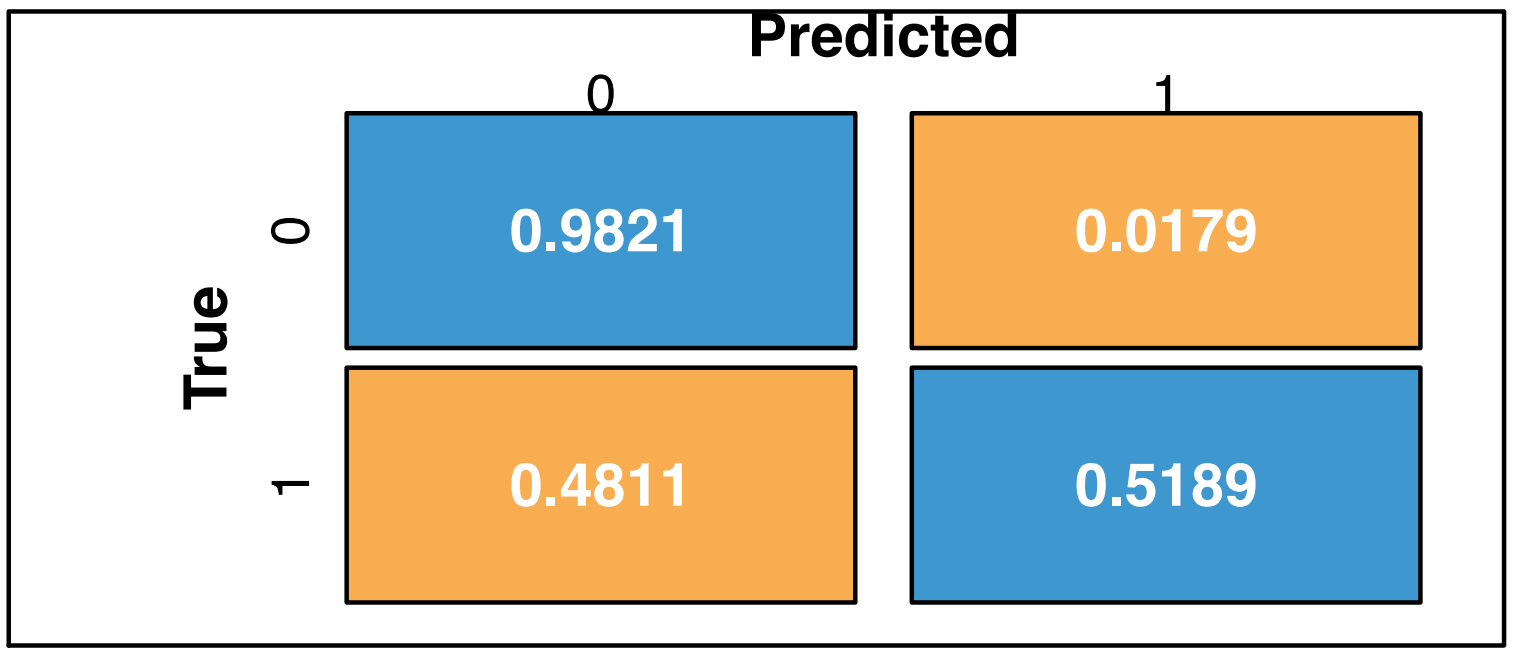

Figure 17: Normalized confusion matrix of LSTM applied to the MHEALTH dataset without oversampling.

```
+     layer_dropout(rate = 0.1) %>%
+     layer_dense(dim(train.y)[2]) %>%
+     layer_dropout(rate = 0.1) %>%
+     layer_activation("softmax")
```

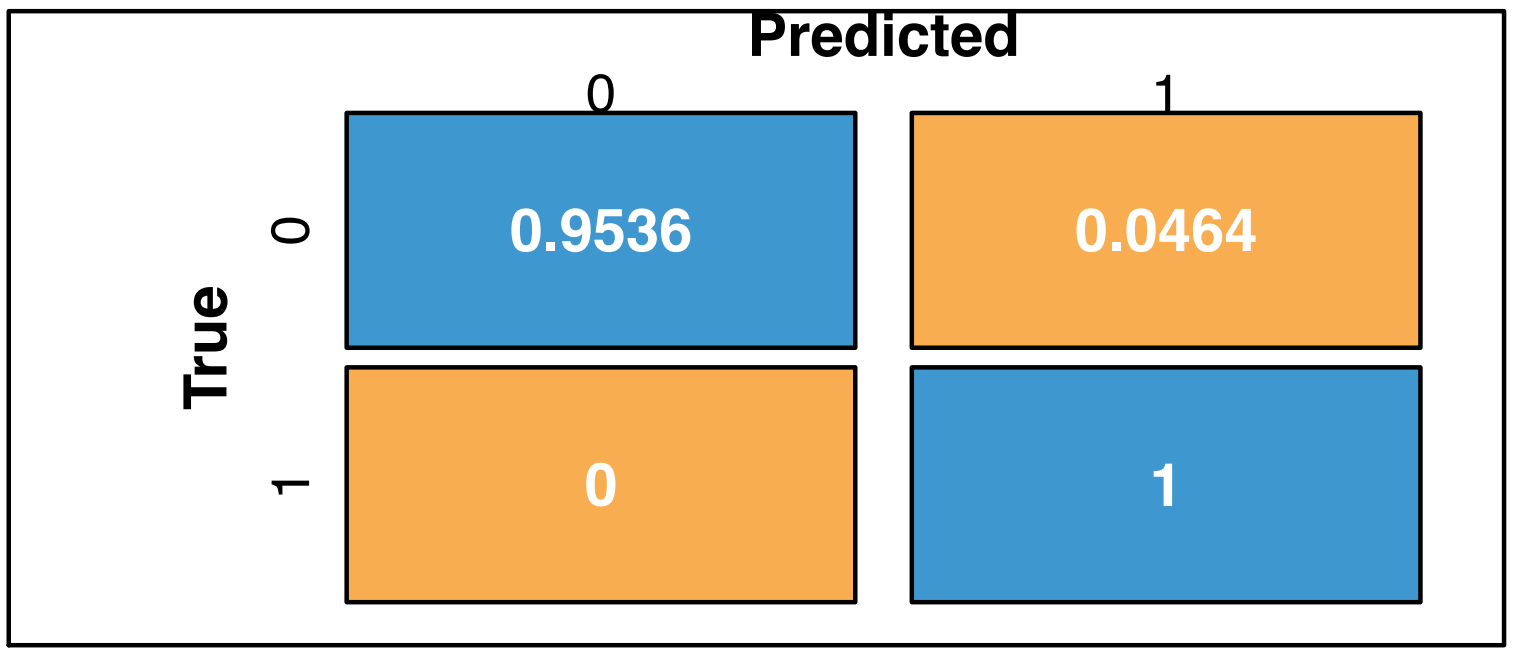

Figure 18: Normalized confusion matrix of LSTM applied to the MHEALTH dataset with oversampling.

```
R> history <- LossHistory$new()
R> model %>% compile(
+    loss = "categorical_crossentropy",
+    optimizer = "adam",
+    metrics = c("accuracy",'f1_score_0' = metric_f1_0, 'f1_score_1' = metric_f1_1,
+                'f1_score_2' = metric_f1_2)
+  )
R> lstm.before <- model %>% fit(
+    x = train.x,
+    y = train.y,
+    validation_data=list(vali.x,vali.y),
+    callbacks = list(history),
+    epochs = 100
+  )

R> model.over <- keras_model_sequential()
R> model.over %>%
+    layer_lstm(10, input_shape = c(dim(train.x)[2], dim(train.x)[3])) %>%
+    layer_dropout(rate = 0.1) %>%
+    layer_dense(dim(train.y)[2]) %>%
+    layer_dropout(rate = 0.1) %>%
+    layer_activation("softmax")
R> history.over <- LossHistory$new()
R> model.over %>% compile(
+    loss = "categorical_crossentropy",
+    optimizer = "adam",
+    metrics = c("accuracy",'f1_score_0' = metric_f1_0, 'f1_score_1' = metric_f1_1,
+                'f1_score_2' = metric_f1_2)
+  )
```

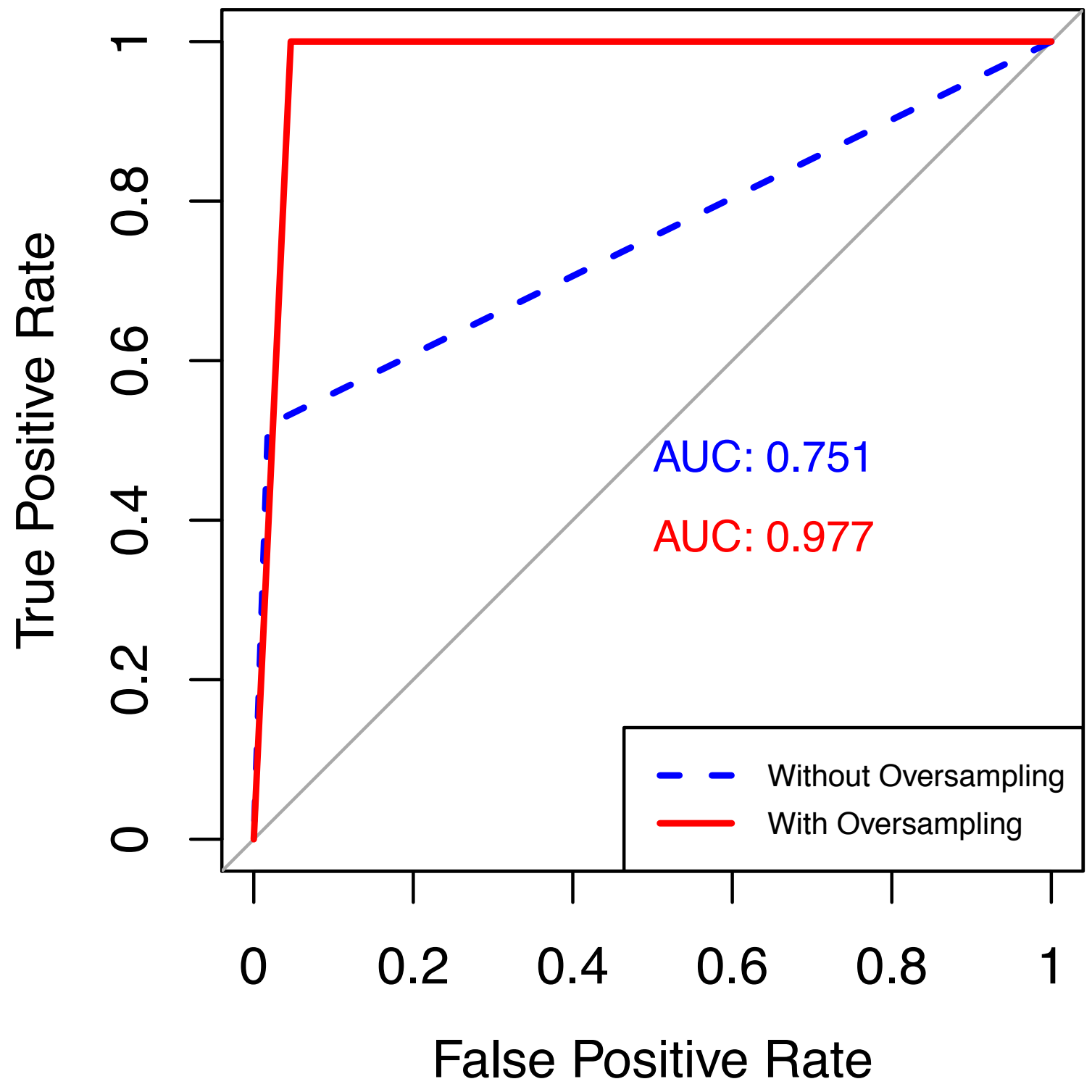

Figure 19: ROC curves of LSTM applied to the MHEALTH dataset, with and without oversampling.

```
R> lstm.after <- model.over %>% fit(
+      x = over.x,
+      y = over.y,
+      validation_data=list(vali.x,vali.y),
+      callbacks = list(history.over),
+      epochs = 100
+  )
```

```
The class 1 F1 score without oversampling:  0.1538462

The class 0 F1 score without oversampling:  0.9757571

The class -1 F1 score without oversampling:  0.1826923

The class 1 F1 score with oversampling:  0.2854311
```

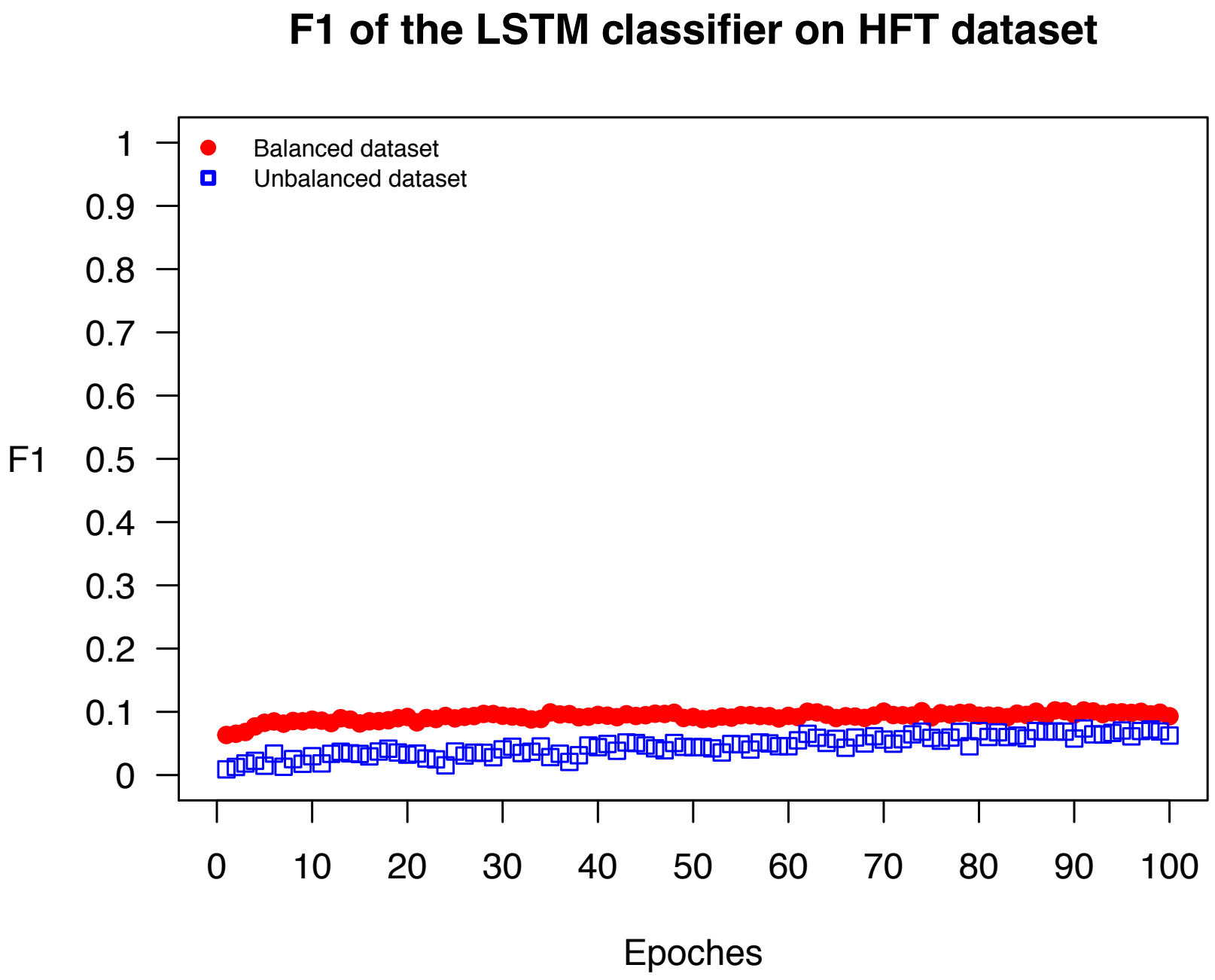

Figure 20: The F1 scores (class 1) of the LSTM classifier trained on the unbalanced and balanced HFT dataset. Both metrics are evaluated at the end of each epoch.

```
The class 0 F1 score with oversampling:  0.9007458

The class -1 F1 score with oversampling:  0.2810127
```

The comparative results are similar to the MHEALTH dataset - oversampling improves the performance and the comparative gain from using `OSTSC()` only increases with more training observations and more epoches.

## 6. Summary

The **OSTSC** package is a powerful oversampling approach for classifying univariant, but multinomial time series data. This article provides a brief overview of the over-sampling methodology implemented by the package. We first provide three examples for the user to verify correct package installation and reproduceability of the results. Using a 'TensorFlow' implementation of an LSTM architecture, we compared the classifier with and without over-sampling. We then repeated the evaluation on two medium size datasets which demonstrate the performance gains from using OSTSC and do not require significant computation. Finally, two large datasets are evaluated to demonstrate the scalability of the package. The examples serve to demonstrate that the OSTSC package improves the performance of RNN classifiers applied to highly imbalanced time series data.

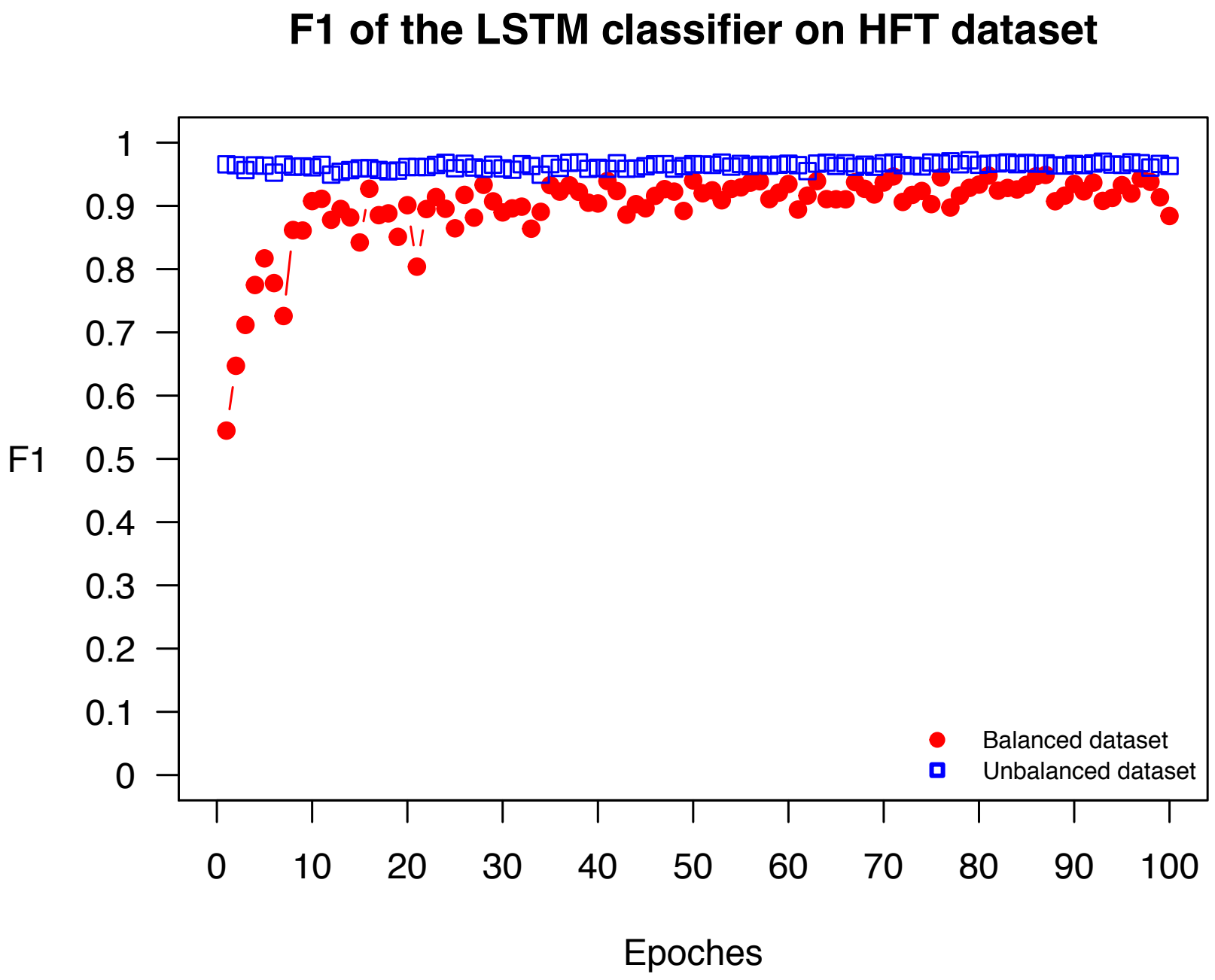

Figure 21: The F1 scores (class 0) of the LSTM classifier trained on the unbalanced and balanced HFT dataset. Both metrics are evaluated at the end of each epoch.

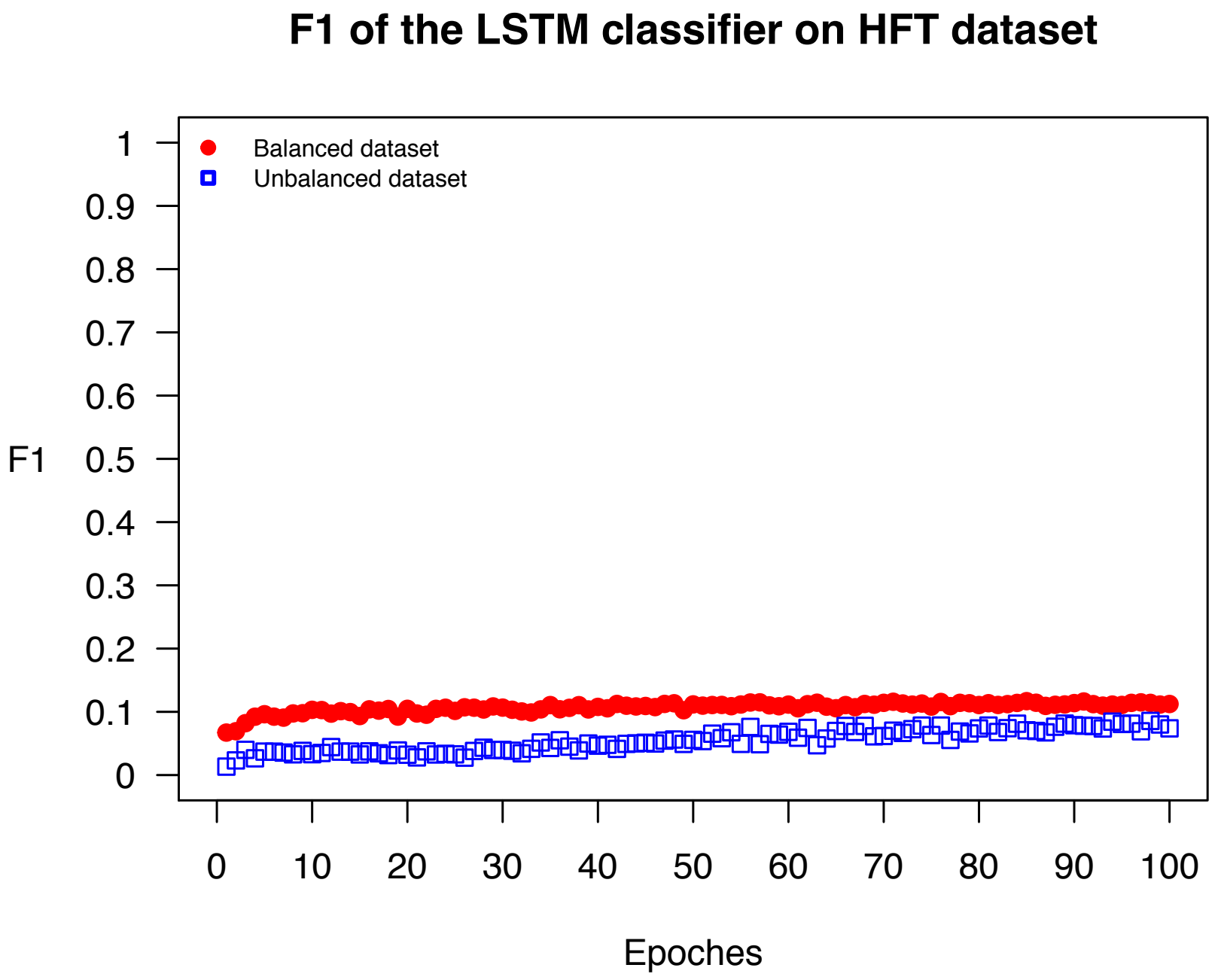

Figure 22: The F1 scores (class -1) of the LSTM classifier trained on the unbalanced and balanced HFT dataset. Both metrics are evaluated at the end of each epoch.

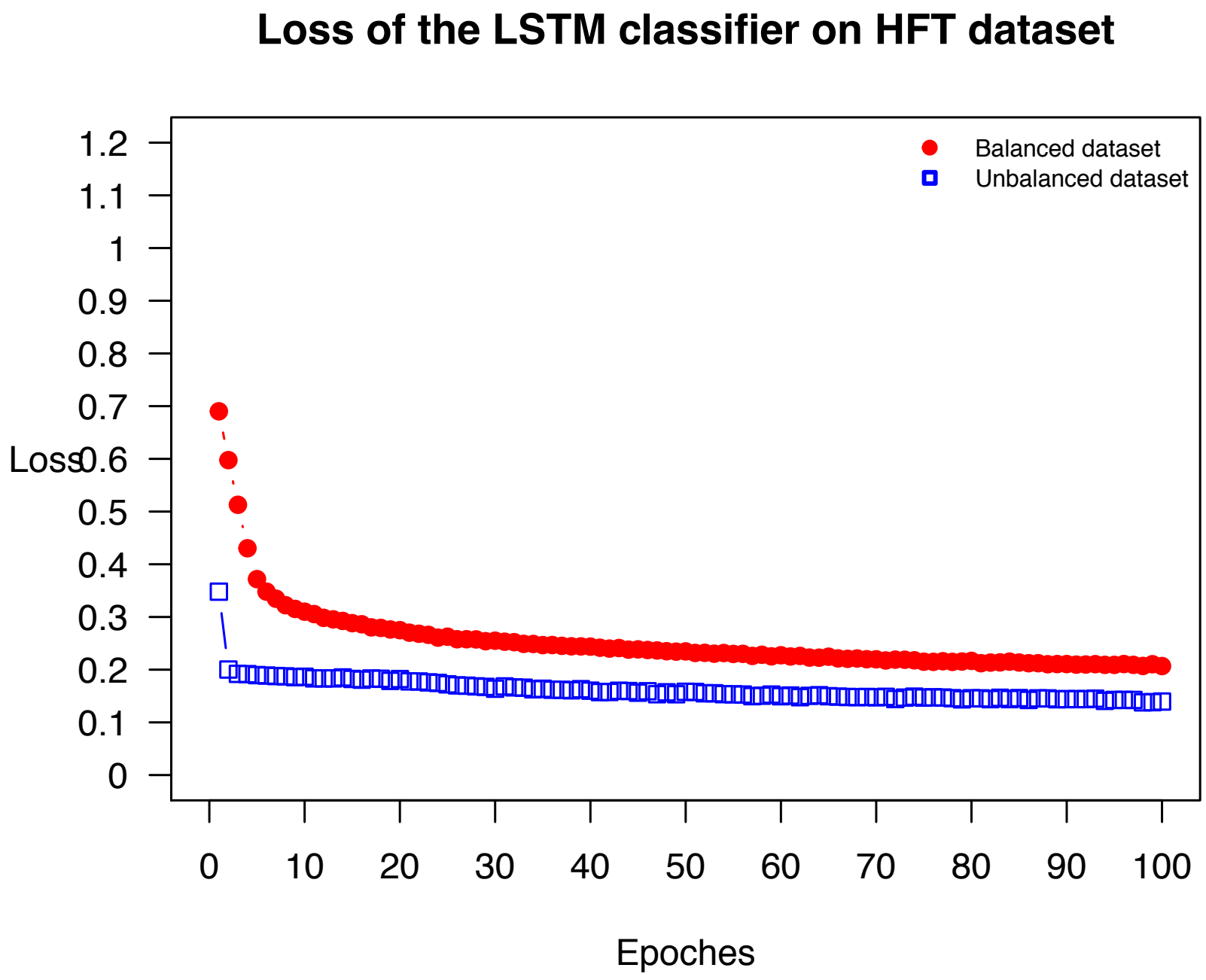

Figure 23: The losses of the LSTM classifier trained on the unbalanced and balanced HFT dataset. Both metrics are evaluated at the end of each epoch.

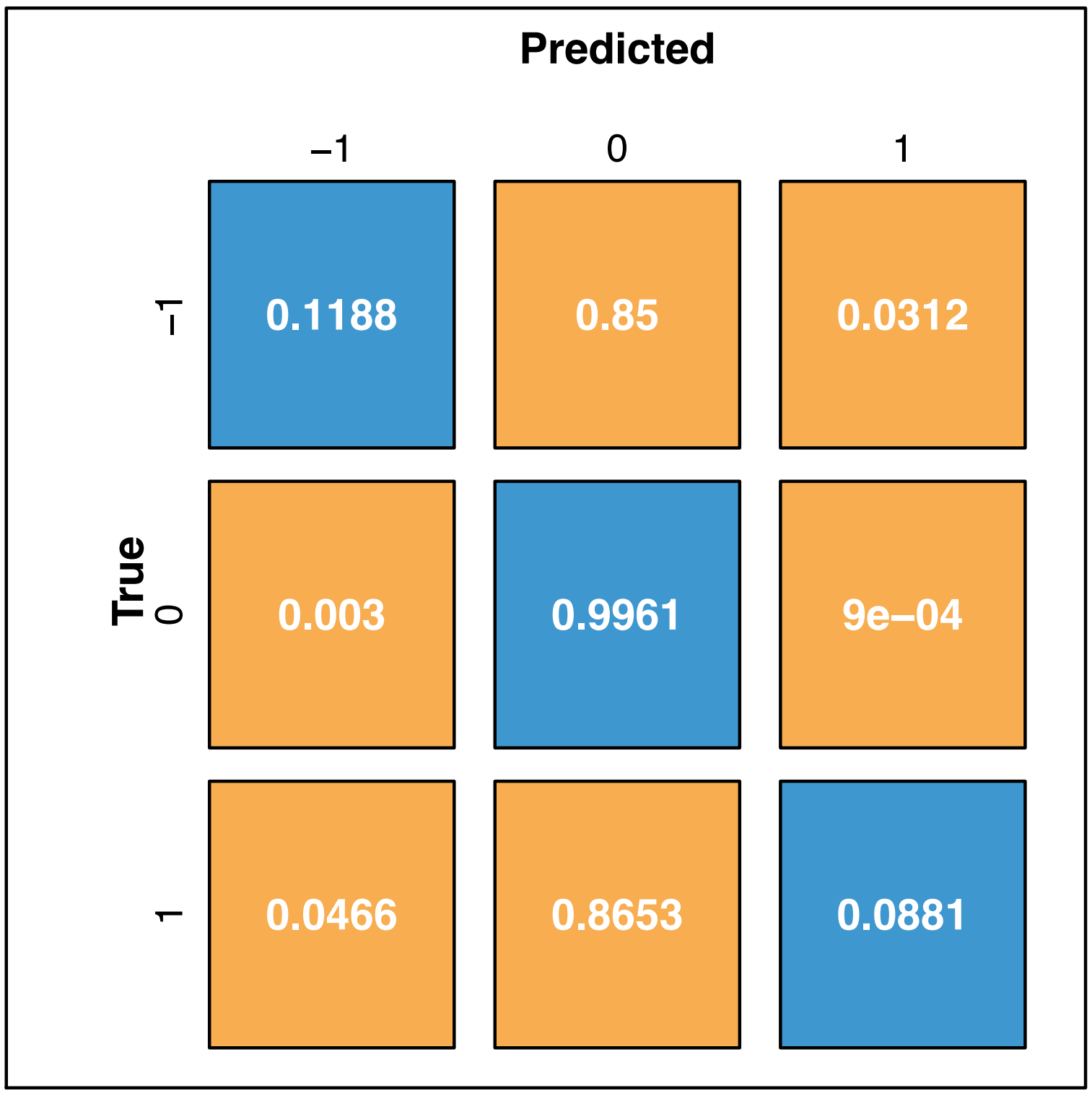

Figure 24: Normalized confusion matrices of LSTM applied to the HFT dataset without oversampling.

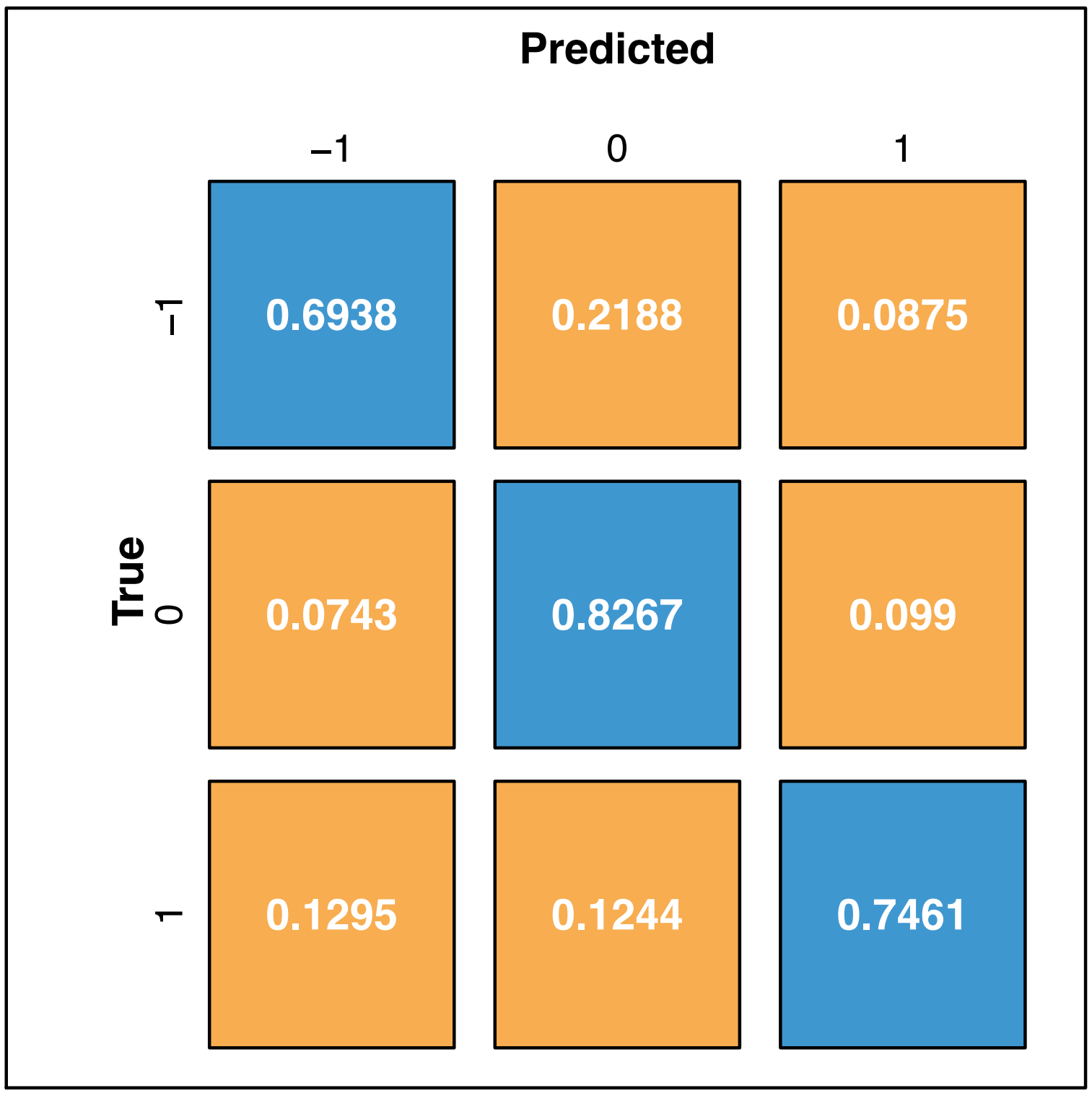

Figure 25: Normalized confusion matrix of LSTM applied to the HFT dataset with oversampling.

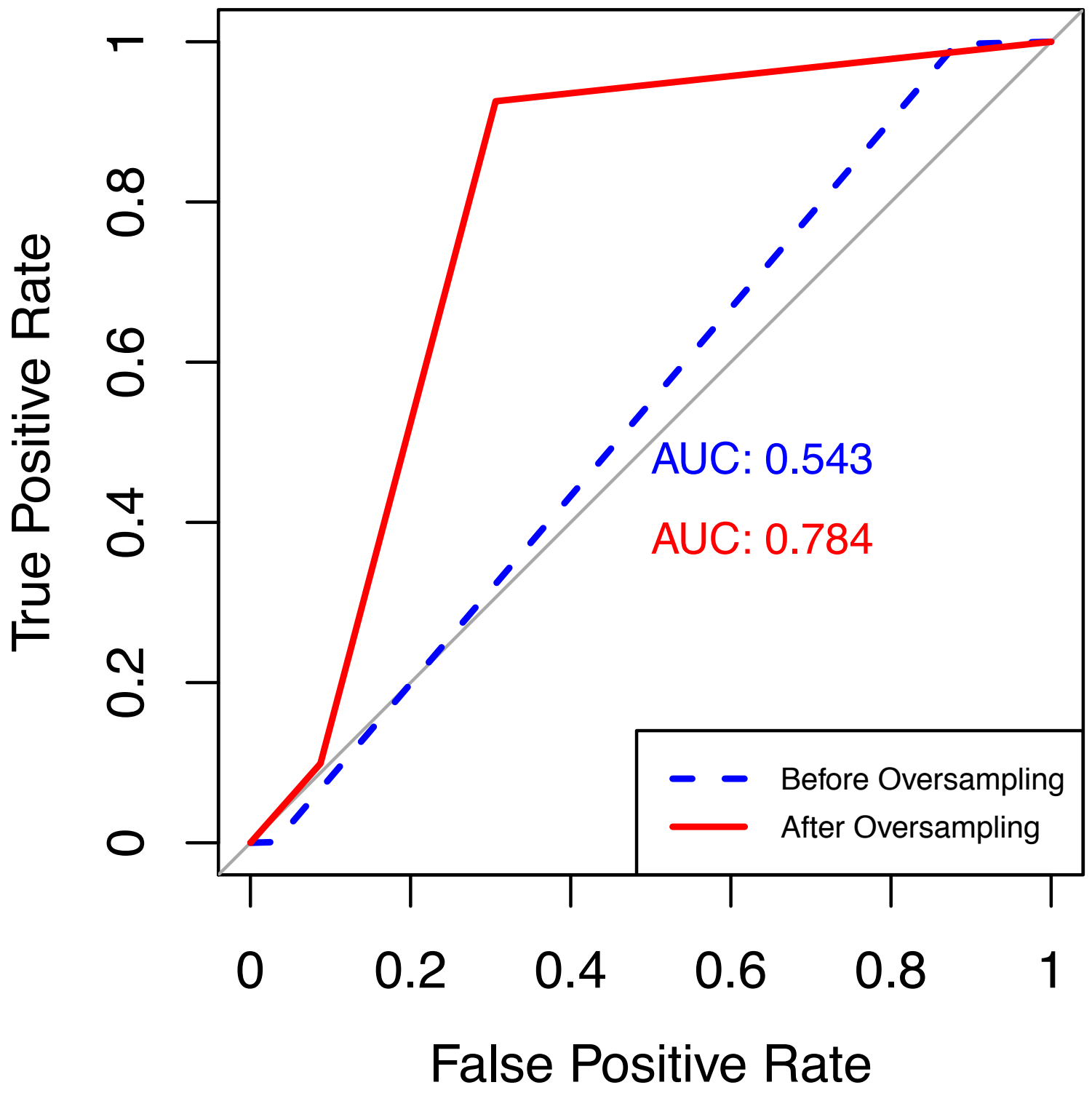

Figure 26: ROC curves of LSTM applied to the HFT dataset with and without oversampling.

**Affiliation:**

Matthew Dixon
Faculty of Statistics and Finance
Stuart School of Business
Illinois Institute of Technology
565 W Adams St, Chicago, IL 60616
E-mail: mdixon7@stuart.iit.edu
URL: https://stuart.iit.edu/faculty/matthew-dixon